\documentclass[aps,prl,reprint,twocolumn,showpacs,floatfix,superscriptaddress]{revtex4-1}

\usepackage{amssymb,amsmath,amstext}                
\usepackage{graphicx}                                               
\usepackage{epstopdf}                                               
\usepackage{color}                                                     
\usepackage{bm}                                                        
\usepackage{appendix}                                              
\usepackage[utf8]{inputenc}
\usepackage{latexsym}
\usepackage{xcolor}
\usepackage{ulem}
\definecolor{lblue} {RGB}{51,71,158}
\usepackage[colorlinks=true,citecolor=blue,linkcolor=blue,urlcolor=lblue]{hyperref}
\DeclareMathOperator{\Tr}{Tr}
\newcommand{\at}[2][]{#1|_{#2}}

\begin{document}

\title{Fidelity susceptibility in Gaussian Random Ensembles
}
\author{Piotr Sierant}
\affiliation{Instytut Fizyki im. Mariana Smoluchowskiego, Uniwersytet Jagiello\'nski,  \L{}ojasiewicza 11, 30-348 Krak\'ow, Poland }
\email{piotr.sierant@uj.edu.pl}

\author{Artur Maksymov}
\affiliation{Instytut Fizyki im. Mariana Smoluchowskiego, Uniwersytet Jagiello\'nski,  \L{}ojasiewicza 11, 30-348 Krak\'ow, Poland }

\author{Marek Ku\'s}
\affiliation{Centrum Fizyki Teoretycznej PAN, Aleja Lotnik\'ow 32/46, 02-668 Warszawa }
\email{marek.kus@cft.edu.pl}

\author{Jakub Zakrzewski}
\affiliation{Instytut Fizyki im. Mariana Smoluchowskiego, Uniwersytet Jagiello\'nski,  \L{}ojasiewicza 11, 30-348 Krak\'ow, Poland }
\affiliation{Mark Kac Complex
Systems Research Center, Uniwersytet Jagiello\'nski, Krak\'ow,
Poland. }
\email{jakub.zakrzewski@uj.edu.pl}

\date{\today}

\begin{abstract}
The fidelity susceptibility measures sensitivity of eigenstates to a change of 
an external parameter. It has been fruitfully used to pin down quantum 
phase transitions when applied to ground states (with extensions to thermal states). 
Here we propose to use the fidelity susceptibility as a useful dimensionless measure 
for complex quantum systems. We find analytically the fidelity susceptibility distributions 
for Gaussian orthogonal and unitary universality classes for arbitrary system size. 
The results are verified by a comparison with numerical data.
\end{abstract}

\maketitle

The discovery of many body localization (MBL) phenomenon resulting in non-ergodicity of the dynamics in many
body systems \cite{Basko06} restored also the interest in purely ergodic phenomena modeled by Gaussian 
random ensembles (GRE) \cite{Mehtabook} and in possible measures to characterize them. 
The gap ratio between adjacent level spacings \cite{Oganesyan07} was introduced precisely for that purpose
as it does not involve the so 
called unfolding \cite{Haake} necessary for meaningful studies of level spacing distributions and yet often leading 
to spurious results \cite {Gomez02}. Still, the level spacing distribution belongs to  the most popular statistical measures  used for single
particle quantum chaos studies \cite{Bohigas84,Bohigas93,Stockmann99,Guhr98} and also in the transition to MBL 
\cite{Serbyn16,Bertrand16,SierantPRB}.
A particular place among different measures was taken by those characterizing level dynamics for a Hamiltonian
$H(\lambda)$ dependent on some parameter $\lambda$. In Pechukas-Yukawa formulation 
\cite{Pechukas83,Yukawa85} energy levels are positions of a fictitious gas particles, derivatives with respect to 
the fictitious time $\lambda$ are velocities (level slopes), the second derivatives describe curvatures 
of the levels (accelerations). Simons and Altschuler \cite{Simons93}  put forward a proposition 
that the variance of velocities distribution is an important parameter characterizing universality 
of level dynamics. This led to predictions for distributions of avoided crossings \cite{Zakrzewski93c} 
and, importantly, curvature distributions postulated first on the basis of numerical data for GRE 
\cite{Zakrzewski93} and then derived analytically \textit{via} supersymmetric
method by von Oppen \cite{vonOppen94,vonOppen95} (for alternative techniques see
\cite{Fyodorov95, Fyodorov11}).  Curvature distributions  were recently addressed
in MBL studies \cite{Filippone16, Monthus17}. 

Apart from  quantum chaos studies in the eighties and nineties of the last millennium, another 
``level dynamics'' tool has been introduced in the quantum information area, i.e. the fidelity, ${\cal F}$ \cite{Uhlmann76}.
It compares two close (possibly mixed) quantum states $\rho(\lambda_1)$ and $\rho(\lambda_2)$ for different values of the parameter, $\lambda$. For pure states, as considered below, and for $\lambda_1=0$ we adopt the following
definition ${\cal F} = |\langle \psi(0) | \psi(\lambda)\rangle|$ \cite{Zanardi06} (note that sometimes
fidelity is defined as a square of ${\cal F}$, such an overlap was considered in the context of parametric dynamics of eigenvectors in \cite{Alhassid95}).
 For sufficiently small difference of parameter values, $\lambda$ it is customary to introduce  a fidelity susceptibility $\chi$ via Taylor series expansion
\begin{equation}
{\cal F}(\hat \rho (0) , \hat \rho(\lambda))= 1 - \frac{1}{2} \chi \lambda^{2} + O(\lambda^{3}),
\label{eq:rozwiniecie}
\end{equation}
(with linear term vanishing due to wavefunction normalization condition).
Fidelity susceptibility is directly related to the quantum Fisher information (QFI), $G$, being
directly proportional to the Bures distance between  density matrices at slightly differing values 
of $\lambda$ \cite{Hubner93, Invernizzi08,Braun18}, with $G(\lambda)=4  \chi $. 

Fidelity susceptibility emerged as a useful tool to study quantum phase transitions as at the 
transition point the ground state changes rapidly leading to the enhancement of
$\chi$ \cite{Zanardi06,You-Li-Gu,Paris,Invernizzi08,Salvatori2014,Bina2016,Kraus16,sanpera_2016}.
All of these studies were restricted to ground state properties while MBL considers the bulk of excited 
states (for a  discussion of thermal states see \cite{Zanardi07,Quan09,Sirker10,Rams18}). 
 In the
context of MBL we are aware of a single study which considered the mean fidelity susceptibility 
across the MBL transition \cite{Hu16}. In particular, nobody addressed the issue of fidelity susceptibility 
behavior for GRE. The aim of this letter is to fill this gap and to provide analytic results for 
the fidelity susceptibility distributions for the most important physically, orthogonal and unitary 
ensembles. This provides novel characteristics of GRE as well as a starting point for the study of 
fidelity susceptibility in the transition to and 
within the MBL domain \cite{Maksymov19}.

Consider $H=H_0+\lambda H_1$ with $H_0, H_1$ corresponding to the orthogonal (unitary) class of GRE
i.e., Gaussian Orthogonal Ensemble (GOE) corresponding to level repulsion parameter $\beta=1$ or Unitary
Ensemble (GUE) with $\beta=2$. For such a Hamiltonian  one may easily prove that fidelity susceptibility of $n$-th eigenstate of $H_0$
is given by
\begin{equation}\label{fidelity}
\chi_n=\sum_{m\ne n}\frac{|H_{1,nm}|^2}{(E_n-E_m)^2},
\end{equation}	
with  $E_n$ being the $n$-th eigenvalue of $H_0$.
We aim at  calculating the probability distribution of the fidelity susceptibility
\begin{equation}\label{probability}
P(\chi,E)=\frac{1}{N\rho(E)}\left\langle \sum_{n=1}^N\delta(\chi-\chi_n)\delta(E-E_n) \right\rangle
\end{equation}
at the energy $E$.
The averaging is over two, independent GRE ($\beta=1,2$)
\begin{equation}\label{pH1}
P(H_a)\sim\exp\left( -\frac{\beta}{4J^2}\Tr{H_a^2}\right), \quad H_a=[H_{a,nm}] 
\end{equation} 	
with $a=0,1$.  Using Fourier representation for $\delta(\chi-\chi_n)$, 
the average over $H_1$ reduces to calculation of Gaussian integrals. Since the formula 
(\ref{fidelity}) involves only the eigenvalues of $H_0$, the averaging over $H_0$ can be expressed as an average over the
well-known joint probability density of eigenvalues \cite{Haake} for a suitable GRE.  At the center of the spectrum ($E=0$),
after straightforward algebraic manipulations (see \cite{Suppl} for details)
we get
\begin{equation}\label{s6}
P(\chi)\sim \int_{-\infty}^{\infty}d\omega 
e^{-i\omega\chi}\left\langle \left[\frac{\det 
\bar{H}^2}{\det\left(\bar{H}^2-\frac{2i\omega J^2}{\beta}
\right)^{\frac{1}{2}}} \right]^\beta \right\rangle_{N-1}, 
\end{equation}
where the averaging is now over $(N-1)\times (N-1)$ matrix $\bar{H}$ from an appropriate Gaussian ensemble. 
Similar averages have been considered in studies
of curvature distributions \cite{vonOppen94,vonOppen95,Fyodorov95},
nonorthogonality effects in weakly open
systems \cite{Poli09, Fyodorov12}
and considered in a more general fashion for the GOE
case in \cite{Fyodorov15}.
\begin{figure}
 \includegraphics[width=1\linewidth]{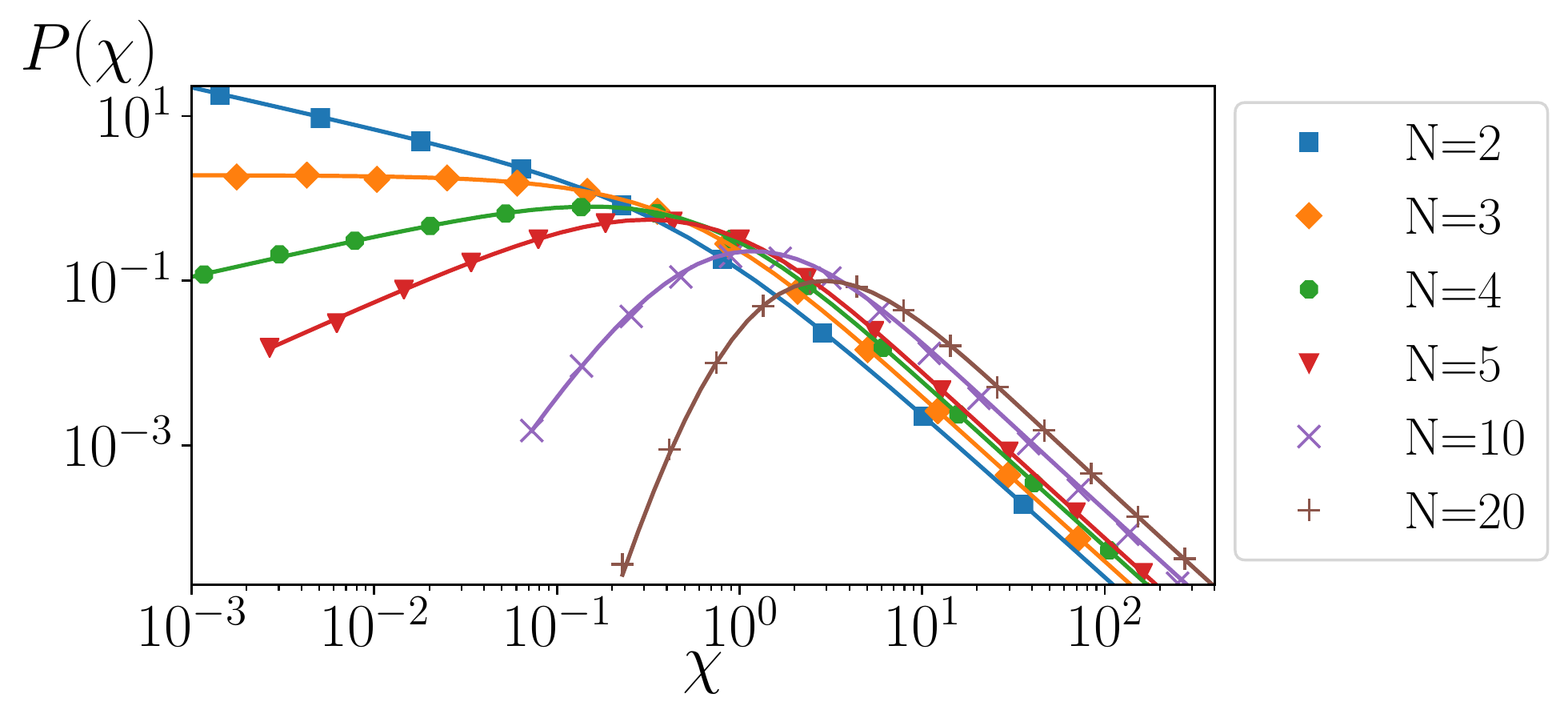}
  \caption{ \label{fig: smallGOE} {Fidelity susceptibility $P^O_N(\chi)$ distribution for GOE matrices of small size $N$. 
  Numerical data denoted by markers.
  Solid lines correspond  to \eqref{eq: 5mt} with $\mathcal{I}^{O,2}_{N}$ given by \eqref{c9}. 
 }}
\end{figure}

To perform the average in \eqref{s6} we employ technique developed in \cite{Fyodorov95}
and express the denominator as a Gaussian integral over 
a vector $\textbf {z} \in \mathbb{R}^{N-1}$ for $\beta=1$ or $\textbf{z}\in \mathbb{C}^{N-1}$ for $\beta=2$.
Employing the invariance of GRE with respect to an adequate class (orthogonal or unitary) of transformations
allows us to choose $\textbf{z}=r[1,0,\ldots,0]^T$, hence we arrive at
\begin{equation}
 P(\chi) \sim \int_{0}^{\infty} dr r^{s}
\delta \left(\chi - 2J^2 r^2/\beta\right) \left \langle \mathrm{det}\bar H^{2\beta} \mathrm{e}^{-r^2 X}
 \right \rangle_{N-1},
 \label{s6a}
\end{equation}
where $X=\sum_{j=1}^{N-1} |\bar H_{1j}|^2$ depends on the first row of $\bar H$ only, and $s = \beta(N-1)-1$. After calculating the ensuing Gaussian integrals over $\bar H_{1j}$ we can reduce the averaging to one over
$(N-2)\times(N-2)$ block of $\bar H$, $V_{ij}=\bar H_{i+1,j+1}$ for $1 \leq i,j \leq N-2$), using the expression 
$\det \bar H=\det V (\bar H_{11}- \sum_{j,k=2}^{N-1} \bar H_{1j} V^{-1}_{jk} \bar H_{1k}^*) $ for a determinant of a block matrix.

Integrating \eqref{s6a} over $r$ we find (details described in \cite{Suppl}) that
the desired fidelity susceptibility distribution $P^O_N(\chi)$ for GOE reads
\begin{eqnarray}
 \nonumber 
 P^O_N(\chi) = 
 \frac{C^{O}_N}{\sqrt{\chi}} \left(\frac{\chi}{1+\chi}
 \right)^{\frac{N-2}{2}}\left( \frac{1}{1+2\chi}\right)^{\frac{1}{2}}\\ 
  \left[ \frac{1}{1+2\chi} + 
 \frac{1}{2} \left(\frac{1}{1+\chi}\right)^{2} 
\mathcal{I}^{O,2}_{N-2}\right],
  \label{eq: 5mt}
\end{eqnarray}
where $C^{O}_N$ is a normalization constant and 
\begin{equation}
 \mathcal{I}^{O,2}_{N}= 
\langle \mathrm{det}V^2\left(2 \mathrm{Tr}V^{-2}+\left(\mathrm{Tr}V^{-1}\right)^2 \right)\rangle_{N}/
 \langle \mathrm{det}V^2\rangle_N.
 \label{5c}
 \end{equation}
 \begin{figure}
 \includegraphics[width=0.52\linewidth]{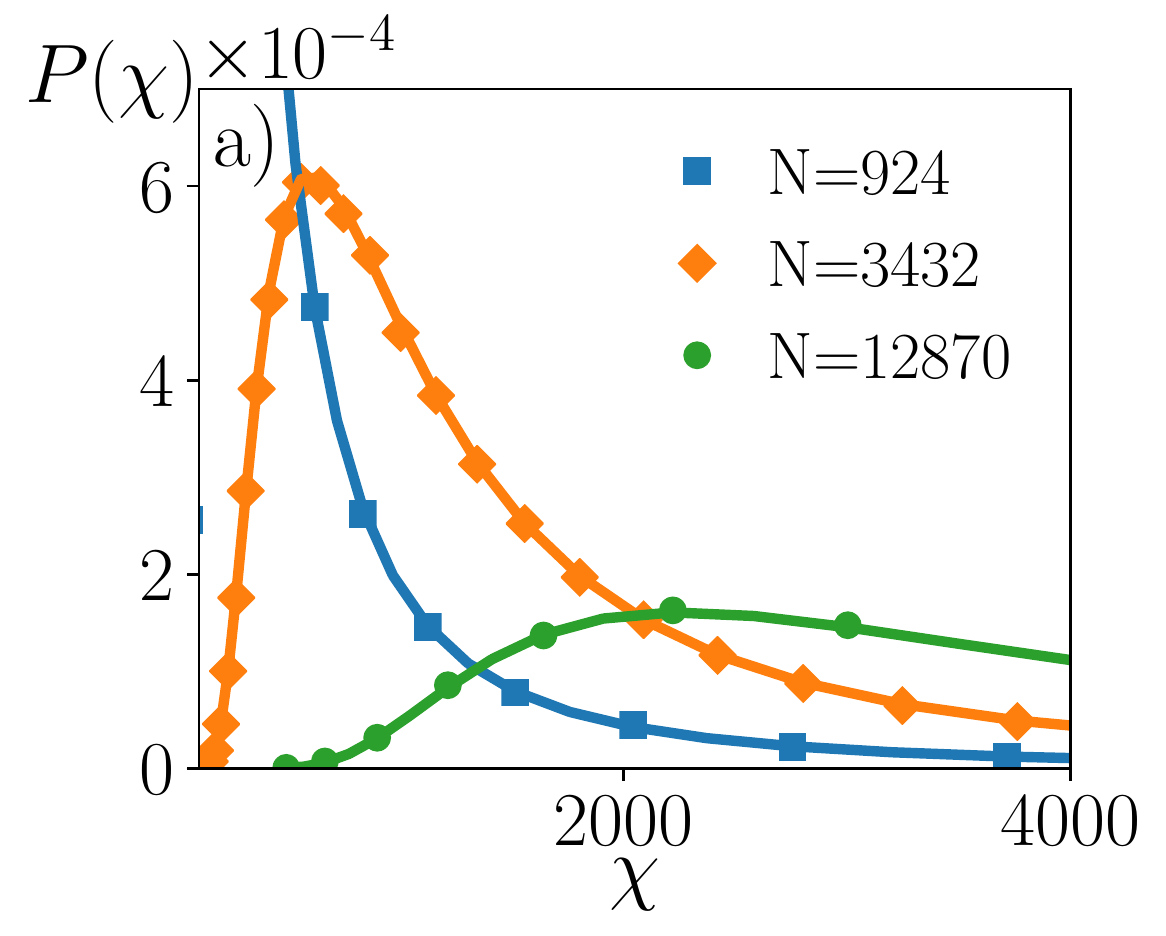}\includegraphics[width=0.5\linewidth]{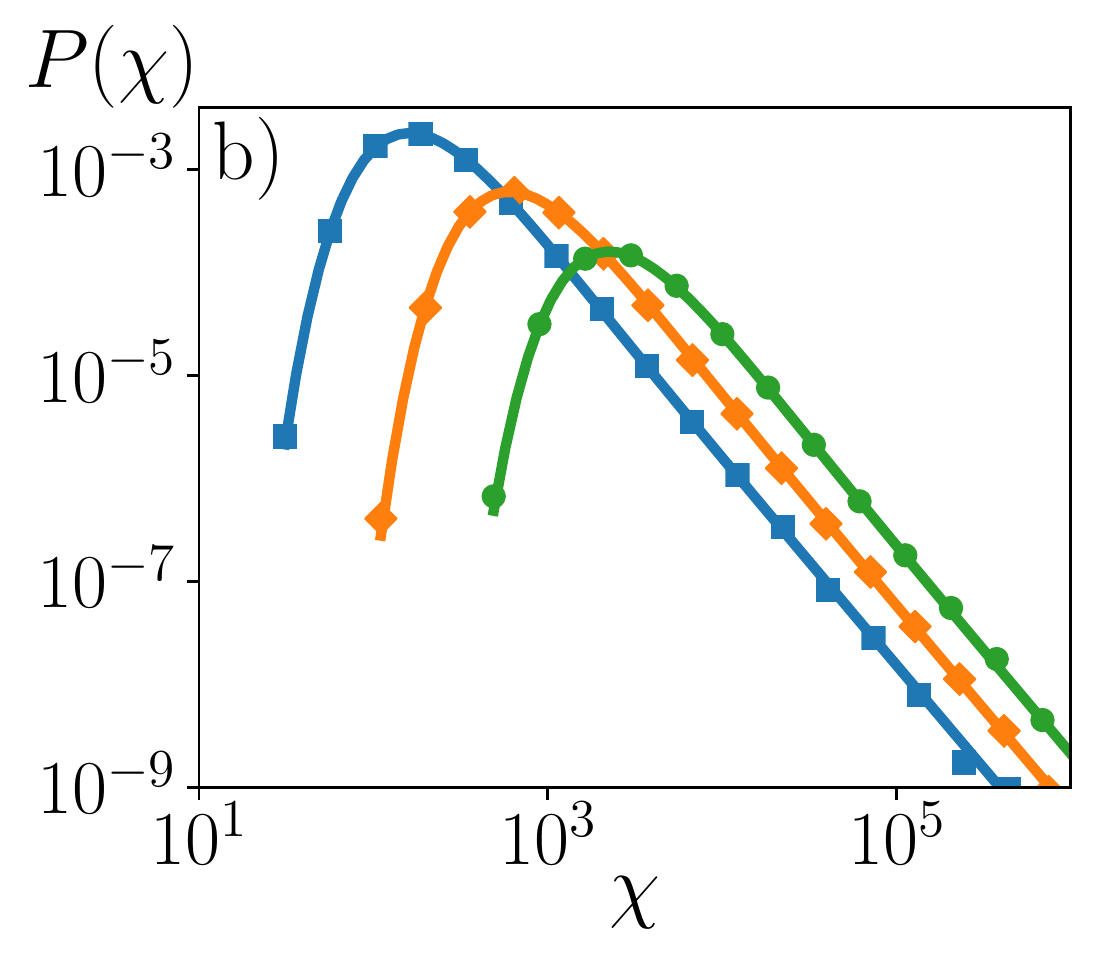}
 \caption{ \label{fig: largeGOE} Fidelity susceptibility $P^O_N(\chi)$ distribution for GOE matrices of
 different sizes as indicated in the Figure. Panels a$)$ and b$)$ correspond to lin-lin
 and log-log scales allowing for a detailed test of accuracy both for the bulk and for the tails 
 of the distribution. Solid lines correspond to \eqref{eq: 5mt} with $\mathcal{I}^{O,2}_{N}$ given by \eqref{c9}.
 }
\end{figure}
\begin{figure}
 \includegraphics[width=0.5\linewidth]{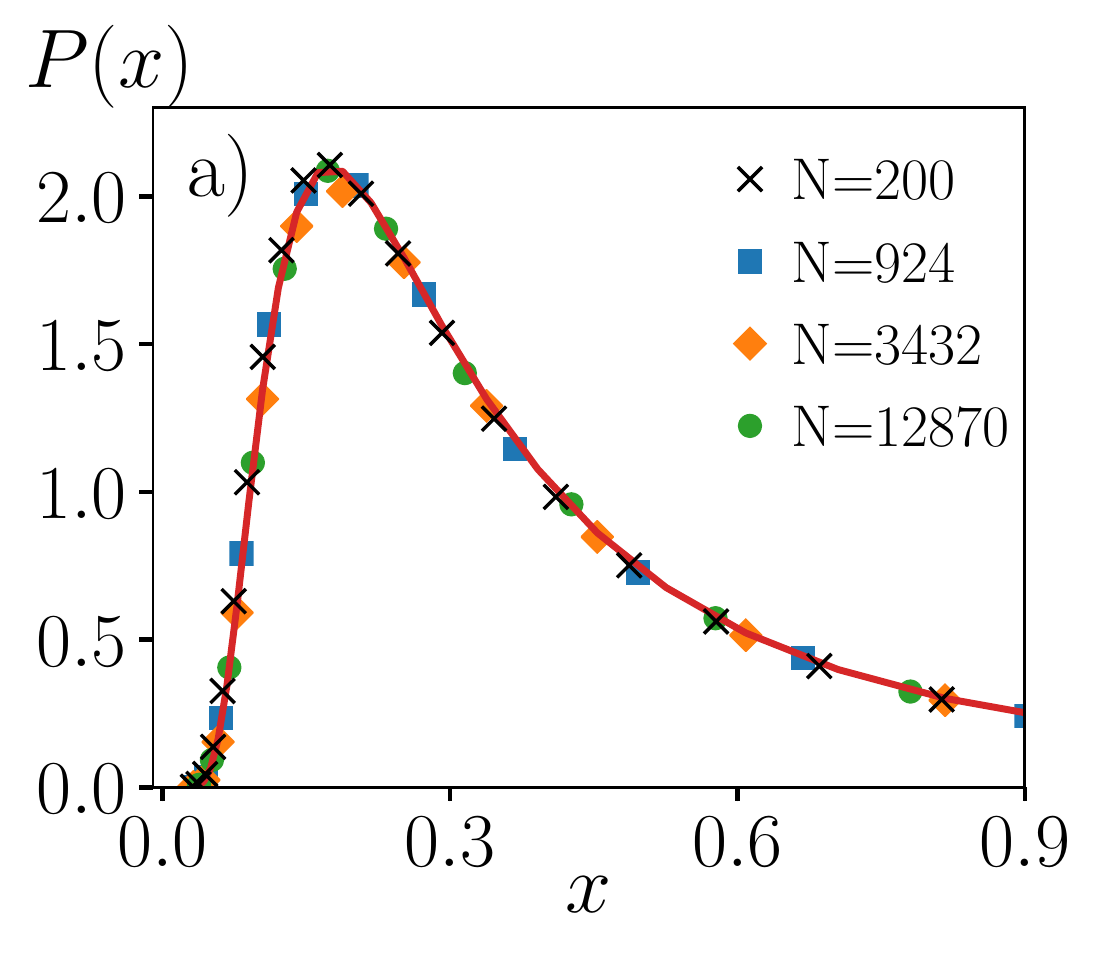}\includegraphics[width=0.51\linewidth]{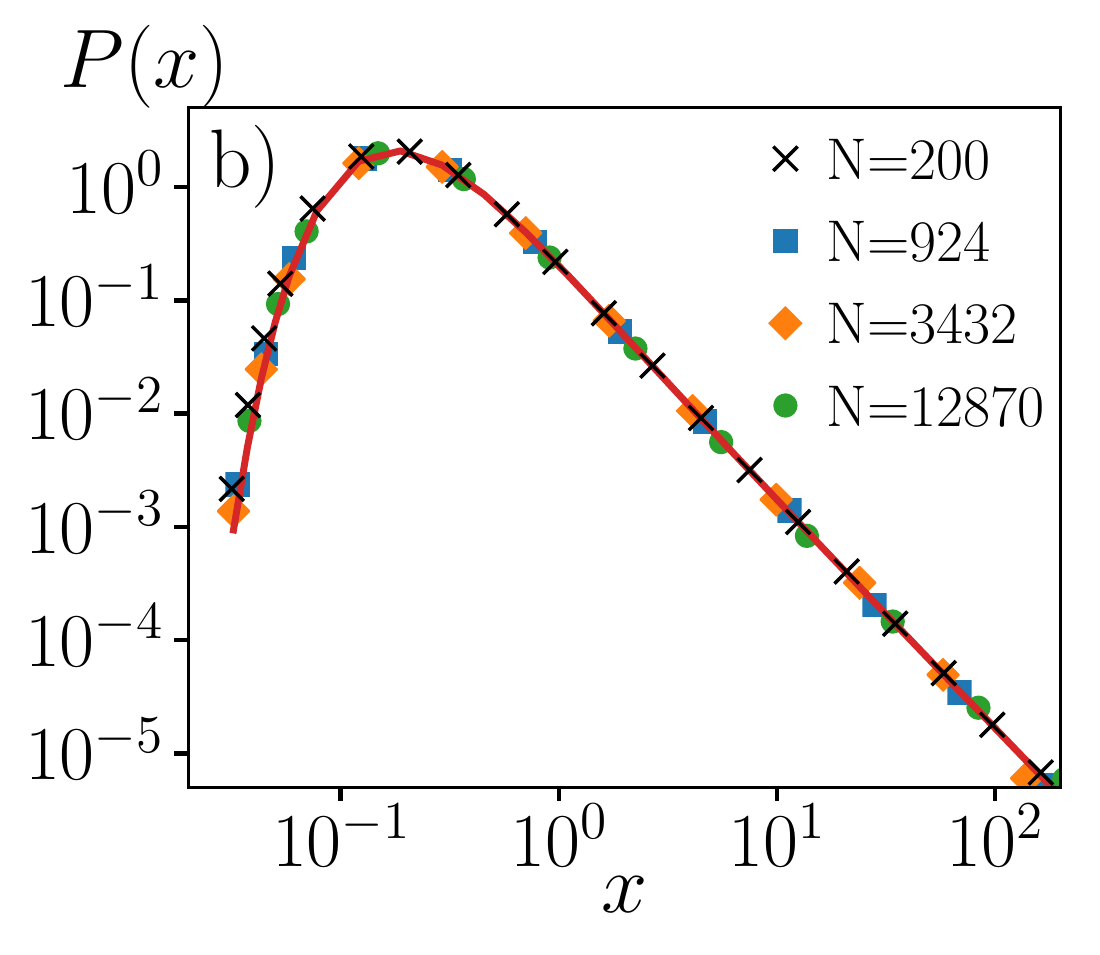}
  \caption{ \label{fig: 2} { Distribution of rescaled fidelity susceptibility 
  $P^O(x)$ for GOE, numerical data denoted by markers, solid lines -- formula \eqref{eq: 7mt}.
 }}
\end{figure}
 The form of \eqref{5c} is suited for a random matrix theory calculation of $ \mathcal{I}^{O,2}_{N}$.
 However,
to obtain $\mathcal{I}^{O,2}_{N}$ it suffices to note that our calculation implies that
\begin{equation}
 \left \langle \mathrm{det}\bar H^{2} \mathrm{e}^{-r^2 X}
 \right \rangle_{N-1} \at[\Big]{r=0}= J^2 \left\langle \det \bar V^2 \right \rangle_{N-2}(  
 \mathcal{I}^{O,2}_{N-2}+2),
 \label{eq: Io2n}
\end{equation}
showing that $\mathcal{I}^{O,2}_{N}$ is actually determined by the second moments
of determinants of matrices of appropriate sizes from GOE.
Moments as well as the full probability distribution of determinant of GOE matrices
were obtained in \cite{Delannay00} for arbitrary $N$. 
Using  the expression for the second moment in \eqref{eq: Io2n} we get 
\begin{equation}
 \mathcal{I}^{O,2}_{N}=
\begin{cases}
N\frac{N+2}{ N+3/2}, \quad  \quad N \quad \mathrm{even}, \\ 
N+1/2,\quad  \quad N \quad \mathrm{odd}.
\end{cases}
 \label{c9}
\end{equation}
The formula \eqref{c9} is exact for arbitrary $N\geq 0$.
Inserting appropriate values of $\mathcal{I}^{O,2}_{N}$ into \eqref{eq: 5mt} we obtain
an exact formula for the fidelity susceptibility distribution $P^O_N(\chi)$ for GOE matrix of arbitrary size $N$.
Comparison of the resulting distribution $P^O_N(\chi)$ with numerically generated fidelity susceptibility
distributions for small matrix sizes $N\leq20$ is shown in Fig.~\ref{fig: smallGOE}. 
However, it is the large $N$ regime which is interesting
from the point of view of potential applications. 
For $N \gg 1$ 
the $\mathcal{I}^{O,2}$ increases linearly $\mathcal{I}^{O,2}_{N}=N$ with the matrix size $N$. This,
together with the form of $P^O_N(\chi)$ implies
that $P^O_{\alpha N}(\alpha \chi) \approx P^O_{N}(\chi)$.
Indeed, the
distribution $P(\chi)$ scales linearly with $N$ as visible in Fig.~\ref{fig: largeGOE}. 
The linear in $N$ scaling of $\chi$ suggests to introduce scaled fidelity susceptibility, $x={\chi}/{N}$.
Inserting it into \eqref{eq: 5mt} and taking $N\rightarrow \infty$ limit one obtains
\begin{equation}
 P^O(x)=\frac{1}{6} \frac{1}{x^2}\left(1+\frac{1}{x}\right) \exp\left(-\frac{1}{2x}\right),
 \label{eq: 7mt}
\end{equation}
which is the final, simple, analytic result for a large size GOE matrix. 
It performs remarkably well also for modest size matrices e.g. $N=200$ -- compare Fig.~\ref{fig: 2}.
For smaller matrices -- for instance for  $N = 20$, the rescaled distribution $P(x)$ 
has a correct large $x$ tail and a nonzero slope at $x=0$ as 
compared to nonanalytic behavior of $P^O(x)$ at $x=0$ in \eqref{eq: 7mt}. 
Observe also that the mean scaled fidelity susceptibility does not exist as the corresponding integral 
diverges logarithmically showing the importance of the heavy tail of the distribution.
The expression \eqref{eq: 7mt} was 
also obtained  in
study of the so called complexness parameter \cite{Poli09}.

Starting from \eqref{s6a} for GUE ($\beta=2$), after a few technical steps (described in detail in \cite{Suppl})
we obtain the following, exact for arbitrary $N$, expression for the fidelity susceptibility distribution:
\begin{widetext}
 \begin{equation}
   P^U_N(\chi)= C^{U}_N \left(\frac{\chi}{1+\chi}\right)^{N-2}
   \left( \frac{1}{1+2\chi}\right)^{\frac{1}{2}} 
 \left[  \frac{3}{4} \left(  \frac{1}{1+2\chi}\right)^{2} + 
 \frac{3}{2} \frac{1}{1+2\chi} \left( \frac{1}{1+\chi} \right)^2  \mathcal{I}^{U,2}_{N-2} + 
 \frac{1}{4}
\left( \frac{1}{1+\chi} \right)^4  \mathcal{I}^{U,4}_{N-2}
 \right],
          \label{eq: 10}
\end{equation}
\end{widetext}
where  $C^{U}_N$ is a normalization constant. 
 $P_U(\chi)$ for GUE  
depends on two $N$-dependent factors $\mathcal{I}^{U,2}_{N-2}$ and $\mathcal{I}^{U,4}_{N-2}$ that remain
to be determined. 
 They take the form \cite{Suppl} 
\begin{equation}
 \mathcal{I}^{U,K}_{N} = 
 \int dZ_{K,N}   \left \langle \mathrm{det}H^4 \left( \sum_{j,k} z_j H^{-1}_{jk} z^*_k 
     \right)^K \right \rangle_{N}
\label{eq: I42a}     
\end{equation}
where $dZ_{K,N} =(\pi J)^K\int \prod_{j}\mathrm{d}^2 z_j \mathrm{e}^{ -\pi \sum_{j}|z_j|^2}/
\left \langle \mathrm{det}H^{4} \right \rangle_{N}$,
  $K=2,4$ and $H$ is $N \times N$ GUE matrix.
  Performing the integration in \eqref{eq: I42a} we find that $\mathcal{I}^{U,2}_N$ can be expressed 
  in the following way
 \begin{equation}
 \mathcal{I}^{U,2}_N = J^2
 \frac{\left \langle \mathrm{det}H^4 \left( \mathrm{Tr}(H^{-2}) + (\mathrm{Tr}H)^{-2}\right) \right \rangle_{N}}
     {\left \langle \mathrm{det}H^{4} 
\right \rangle_{N}}.
\label{eq: I42ap}
\end{equation}
    Introducing the following generating function
 \begin{equation}
 Z_N(j_1,j_2) = \left \langle \det H^2 \det(H-j_1) \det(H-j_2) \right \rangle_{N},
 \label{eq: 7.1mt}
\end{equation}
we immediately verify that 
\begin{equation}
 \mathcal{I}^{U,2}_N =
     \frac{ J^2}{ Z_N(0,0)
} \left ( 
 2 \frac{\partial^2}{\partial j_1 \partial j_2} Z_N(0,0)
 -\frac{\partial^2}{\partial j_1^2 } Z_N(0,0)
\right ).
\label{eq: 7.5mt}
\end{equation}
 The generating function $Z_N(j_1,j_2)$ is actually a
 correlation function of a characteristic polynomial of the $H$ matrix.
 It was shown in \cite{Brezin00, Fyodorov03} that such quantities can be calculated exactly 
 as determinants of appropriate orthogonal polynomials.
 A kernel structure of those expressions has been identified in
 \cite{Strahov03} leading to formulas most convenient in our calculation of $Z_N(j_1,j_2)$.
 The generating function $Z(j_1,j_2)$ is given by 
 \begin{eqnarray}
\nonumber Z_N(j_1, j_2) & =  &
\frac{C_{N,2}}{(j_1-j_2)} \lim_{\mu_2 \rightarrow 0}\frac{\partial}{\partial \mu_2} \\
& \det &\left[
\begin{array}{cc}
W_{N+2}(j_1,0) & W_{N+2}(j_2, 0) \\
 W_{N+2}(j_1,\mu_2) & W_{N+2}(j_2,\mu_2) \\
\end{array}
\right],
\label{eq: 8.3}
\end{eqnarray}
 with the kernel $W_{N+2}(\lambda, \mu)$ defined as
 \begin{equation}
W_{N+2}(\lambda, \mu) 
=
\frac{ H_{N+2}(\lambda) H_{N+1}(\mu) - 
H_{N+2}(\mu) H_{N+1}(\lambda)}{\lambda-\mu}. 
\label{eq: 8.2} 
\end{equation}
The Hermite polynomials $H_N(\lambda)$
are orthogonal with respect to the measure $\mathrm{e}^{-\frac{1}{2J^2} x^2} \mathrm{d}x$ 
and normalized in such a way that the coefficient in front of $\lambda^N$ is equal to unity.
 \begin{figure}
 \includegraphics[width=0.5\linewidth]{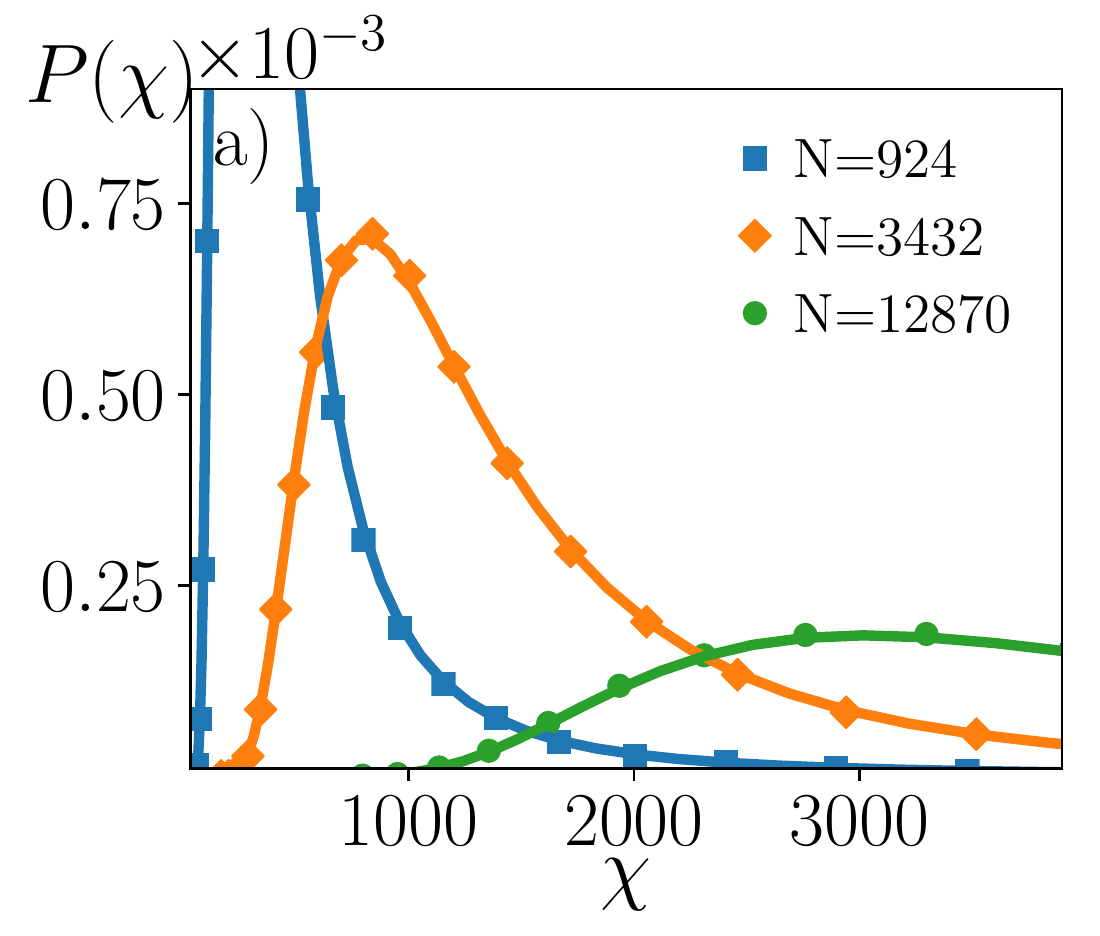}\includegraphics[width=0.5\linewidth]{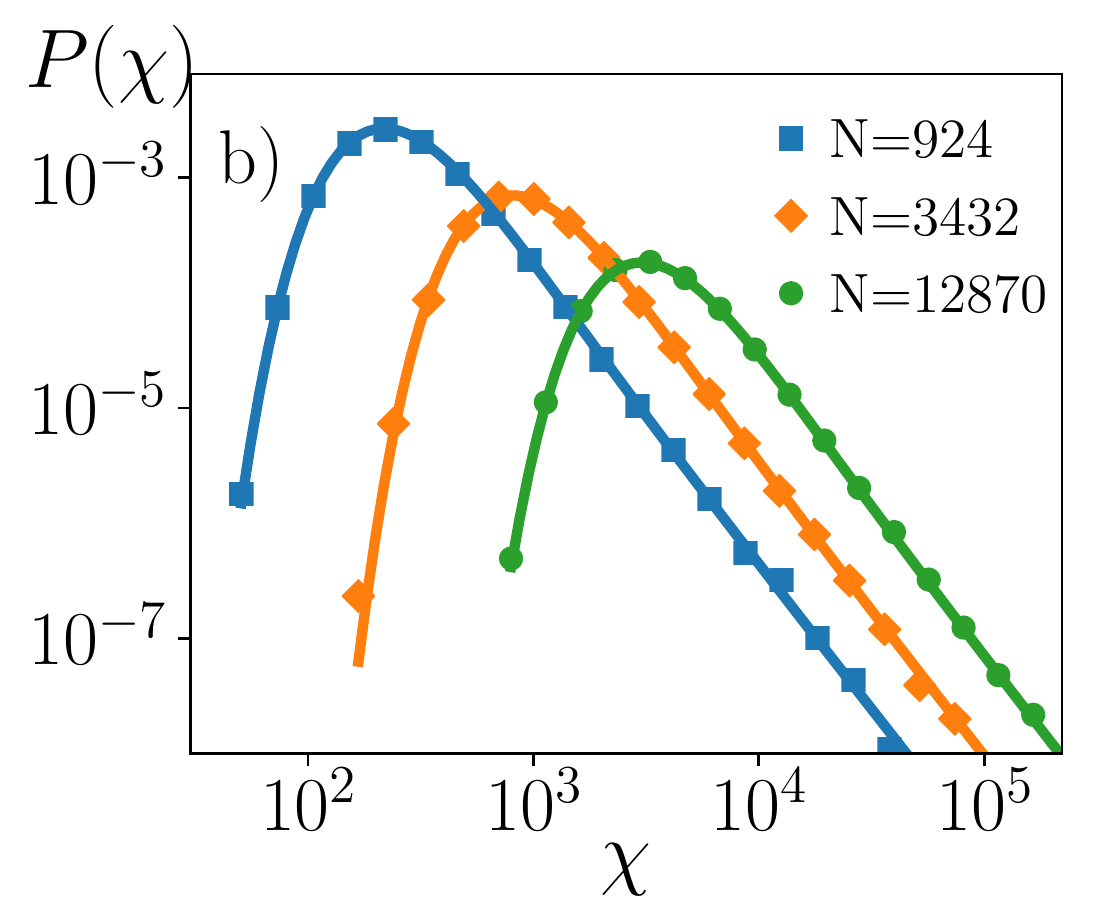}
  \caption{ \label{fig: 3a} Fidelity susceptibility distribution $P^U_N(\chi)$ for GUE, numerically 
  generated data denoted by markers, solid lines -- formula \eqref{eq: 10} with $\mathcal{I}^{U,2}_N $
  and $\mathcal{I}^{U,4}_N $ given by \eqref{eq: 8.9mt} and \eqref{iu4a} respectively.
 }
\end{figure}
\begin{figure}
 \includegraphics[width=0.5\linewidth]{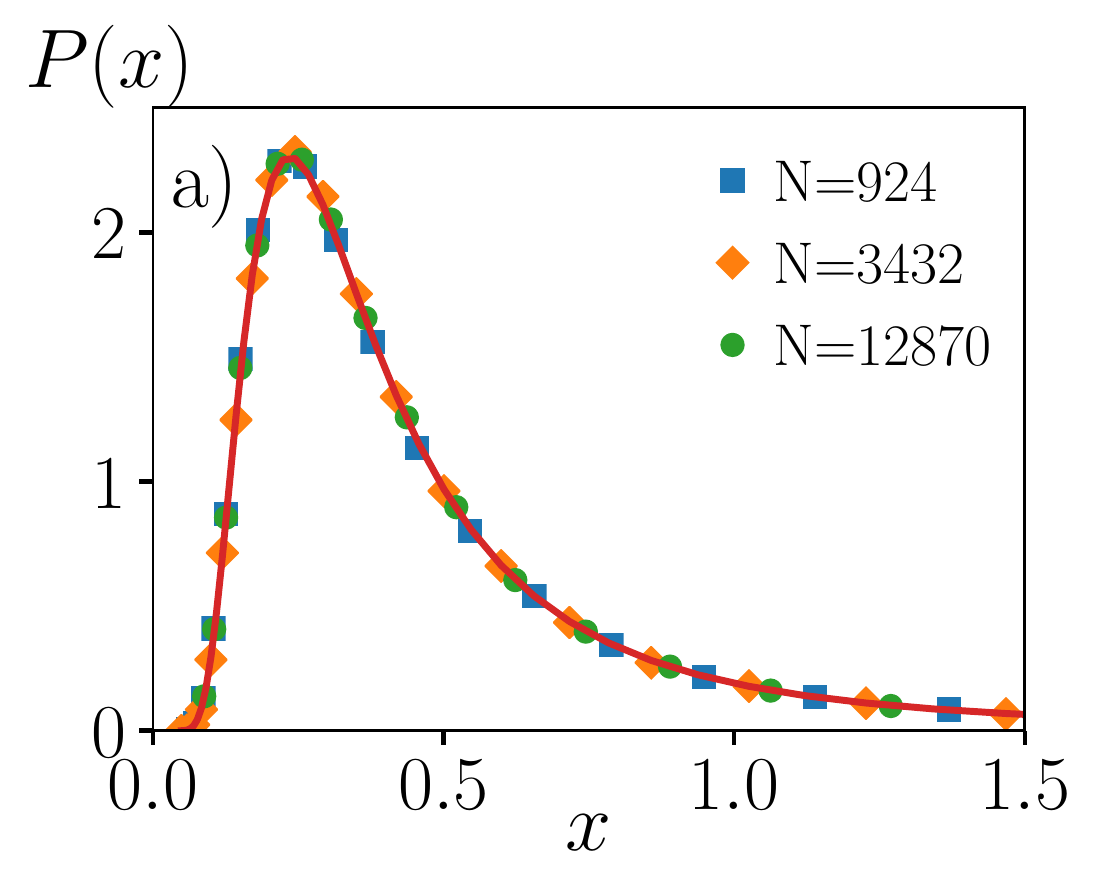}\includegraphics[width=0.5\linewidth]{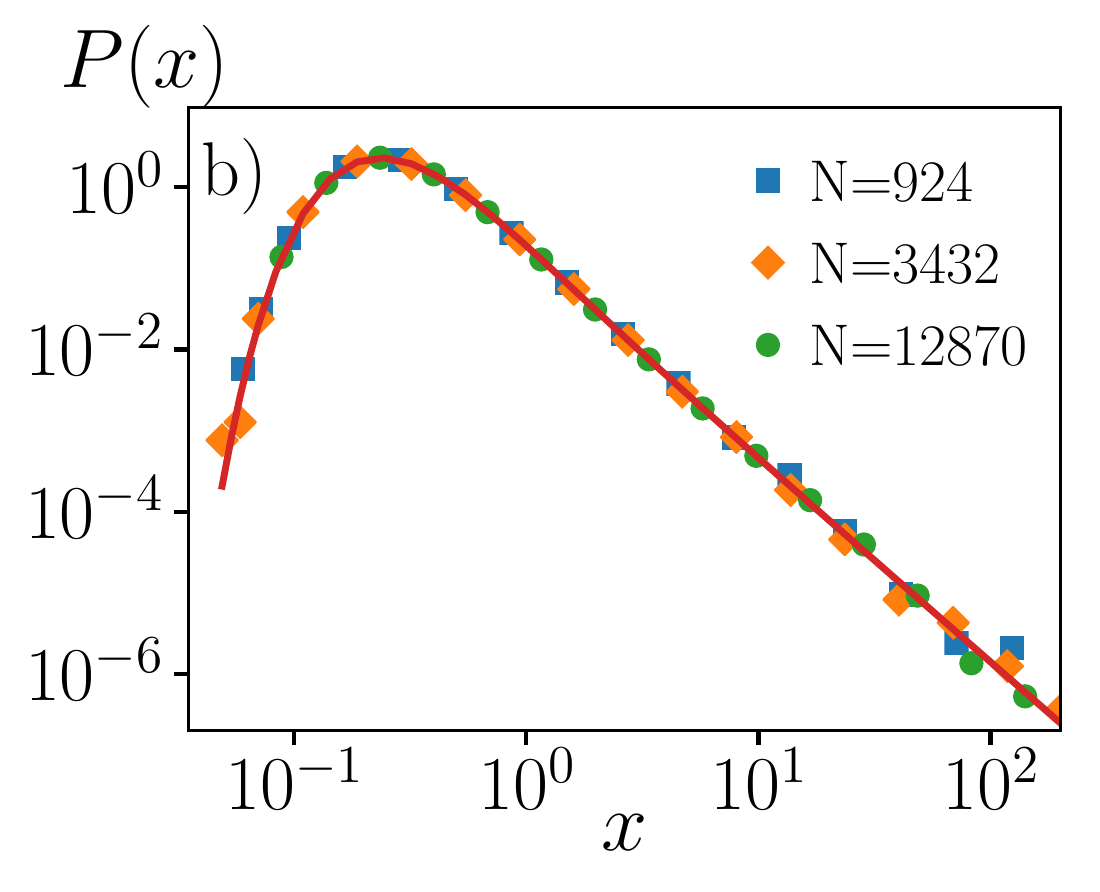}
  \caption{ \label{fig: 3} { 
  Distribution of rescaled fidelity susceptibility 
  $P_U(x)$ for GUE, numerical data denoted by markers, solid lines -- formula \eqref{eq: 11mt}.
 }}
\end{figure}
We have found a closed formula
for the generating function $Z_N(j_1,j_2)$ (see \cite{Suppl} for details).
Calculating the derivatives 
in \eqref{eq: 7.5mt} and taking the limits
$j_1 \rightarrow 0$ and $j_2 \rightarrow 0$ we obtain
\begin{equation}
 \mathcal{I}^{U,2}_N=
\begin{cases}
\frac{1}{3}N, \quad  \quad \quad \quad \,\,\, N \quad \mathrm{even}, \\ 
\frac{1}{3}(N+1), \quad  \quad N \quad \mathrm{odd}.
\end{cases}
 \label{eq: 8.9mt}
\end{equation}
The next step is to use 
the idea analogous to 
the argument with ratio of second moments of determinants of GOE matrices
which allowed us to obtain the exact expression for $\mathcal{I}^{O,2}_{N}$ \eqref{eq: Io2n}.
Employing formulas for the fourth moment of determinant of GUE matrix  \cite{Mehta98, Cicuta00}
and taking into account the expression for $\mathcal{I}^{U,2}_N$ we obtain 
\begin{equation}
 \mathcal{I}^{U,4}_{N} =\begin{cases}
N^2+2N, \quad  \quad \quad \,\,\,\,\, N \quad \mathrm{even}, \\ 
N^2+4N+3, \quad  \quad N \quad \mathrm{odd}.
\end{cases}
 \label{iu4a}
 \end{equation}
 The distribution \eqref{eq: 10} together with expressions \eqref{eq: 8.9mt}, \eqref{iu4a}
 for $\mathcal{I}^{U,4}_N$ and $\mathcal{I}^{U,2}_N$ is the exact fidelity susceptibility
 distribution for GUE for arbitrary $N$.
As shown in Fig.~\ref{fig: 3a} the expression \eqref{eq: 10} is confirmed by numerical
data for different system sizes $N \gg 1$. Similar, perfect agreement of our
formula $P^U_N(\chi)$ with numerically generated data is obtained for small $N \geq 2$  (data not shown).
Moreover, similarly to the GOE case, $P^U_N(\chi)$
scales linearly with increasing 
$N$. Therefore, considering again the distribution of scaled fidelity susceptibility $ x=\chi/N$
we arrive at the large $N$ limit of the simple form 
\begin{equation}
 P^U(x)=\frac{1}{3\sqrt{\pi}}\frac{1}{x^{5/2}}\left(\frac{3}{4}+\frac{1}{x} 
 +\frac{1}{x^2}\right) \exp\left(-\frac{1}{x}\right)
 \label{eq: 11mt}
\end{equation}
which works well for GUE data as shown in Fig.~\ref{fig: 3}.

\begin{figure}
 \includegraphics[width=0.5\linewidth]{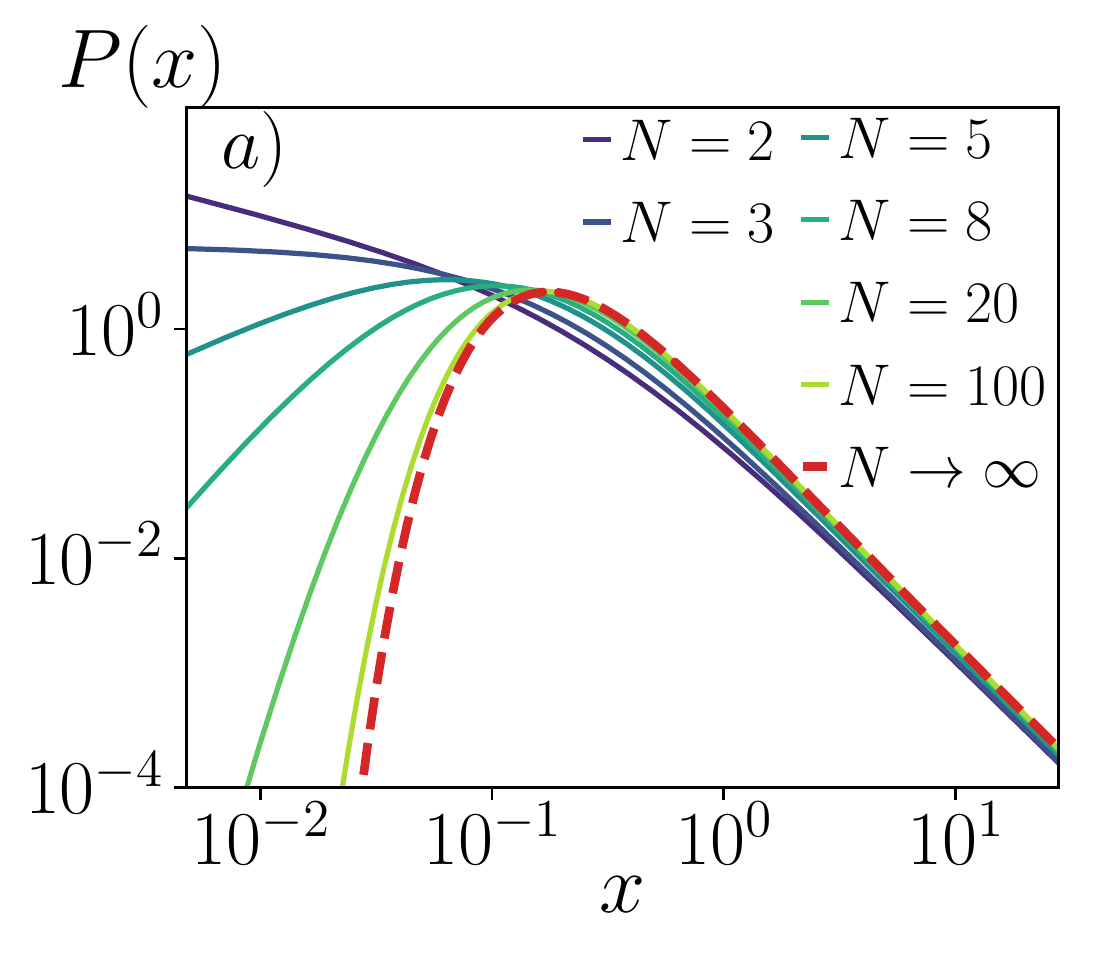}\includegraphics[width=0.5\linewidth]{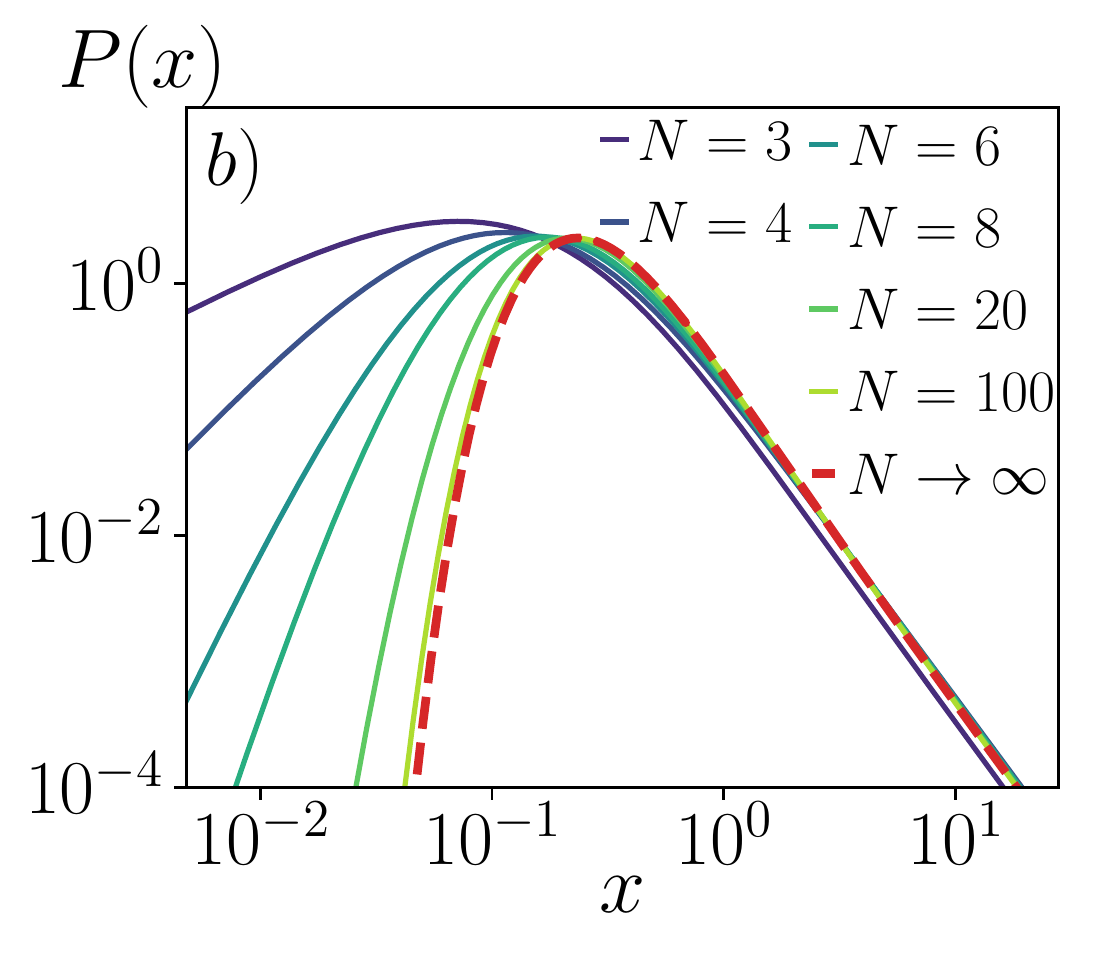}
  \caption{ \label{fig: 6} { 
  Distributions of rescaled fidelity susceptibility $P(x)$ for a) GOE  and b) GUE for matrix size $N$.
 Red dashed lines correspond to the universal $N\rightarrow \infty$ GOE/GUE limits.
 }}
\end{figure}

Remarkably, the obtained distributions of fidelity susceptibility
both for GOE \eqref{eq: 5mt} and GUE \eqref{eq: 10} are exact for arbitrary $N\geq2$. This is unusual situation, 
even for GRE -- for instance, the simple analytic form of the level spacing distribution $P(s)$  
for $N=2$ becomes more complicated for larger $N$ \cite{Mehtabook}. We study thus the 
onset of universal large $N$ behavior of the rescaled fidelity susceptibility distribution $P(x)$. 
The results are shown in Fig.~\ref{fig: 6}. Clearly, the power-law tail of the distributions 
for GOE (GUE) is observed for all $N$. This power-law tail arises in instances when the sum for $\chi_n$ 
\eqref{fidelity} is dominated by a single term with small energy denominator. The algebraic decay
$x^{-2}$ $(x^{-5/2})$ for GOE (GUE) can be derived from the small $s$ behavior of level spacing distribution 
$P(s)$ \cite{Gaspard90, Monthus17}. The approach to the limiting $N \rightarrow \infty$ distributions 
$P^{O,U}(x)$ is associated with decreasing number of instances of very small fidelity susceptibility.

To conclude, we have derived closed formulae for fidelity susceptibility distributions 
corresponding to level dynamics for  both the orthogonal and the unitary class of 
Gaussian random ensembles. Particularly simple analytic expressions are found  in 
the large $N$ limit.
The fidelity susceptibility distributions obtained for quantally chaotic 
systems may be compared with the results found for GOE (GUE) in order to
characterize the degree to which a given system is faithful to random matrix predictions. 
The obtained distributions  also open a way to address level dynamics in the transition between delocalized -- 
ergodic and many-body localized regimes \cite{Maksymov19}.

As a last touch let us mention that fidelity susceptibility
is experimentally accessible by Bragg
spectroscopy \cite{Shi-Jian14}, e.g. in ultra-cold atomic systems \cite{Ernst10,Inguscio}
or by a direct  measurement of many-body wave functions overlap either in a NMR
setting \cite{Zhang2008} or  a system of ultra-cold bosons \cite{Islam15}. 
That paves a way for comparing experimental measurements 
with universal features of fidelity susceptibility distribution
provided in this work.

\begin{acknowledgments} 
We thank Dominique Delande for careful reading of this manuscript.
P. S. and J. Z. acknowledge support by PL-Grid Infrastructure and EU project the EU H2020-FETPROACT-2014 Project QUIC No.641122. 
This research has been supported by 
National Science Centre (Poland) under projects  2015/19/B/ST2/01028 (P.S. and A.M.), 2018/28/T/ST2/00401 
(doctoral scholarship -- P.S.), 2017/25/Z/ST2/03029 (J.Z.), and 2017/27/B/ST2/0295 (M.K.). 
\end{acknowledgments}


\begin{thebibliography}{57}%
\makeatletter
\providecommand \@ifxundefined [1]{%
 \@ifx{#1\undefined}
}%
\providecommand \@ifnum [1]{%
 \ifnum #1\expandafter \@firstoftwo
 \else \expandafter \@secondoftwo
 \fi
}%
\providecommand \@ifx [1]{%
 \ifx #1\expandafter \@firstoftwo
 \else \expandafter \@secondoftwo
 \fi
}%
\providecommand \natexlab [1]{#1}%
\providecommand \enquote  [1]{``#1''}%
\providecommand \bibnamefont  [1]{#1}%
\providecommand \bibfnamefont [1]{#1}%
\providecommand \citenamefont [1]{#1}%
\providecommand \href@noop [0]{\@secondoftwo}%
\providecommand \href [0]{\begingroup \@sanitize@url \@href}%
\providecommand \@href[1]{\@@startlink{#1}\@@href}%
\providecommand \@@href[1]{\endgroup#1\@@endlink}%
\providecommand \@sanitize@url [0]{\catcode `\\12\catcode `\$12\catcode
  `\&12\catcode `\#12\catcode `\^12\catcode `\_12\catcode `\%12\relax}%
\providecommand \@@startlink[1]{}%
\providecommand \@@endlink[0]{}%
\providecommand \url  [0]{\begingroup\@sanitize@url \@url }%
\providecommand \@url [1]{\endgroup\@href {#1}{\urlprefix }}%
\providecommand \urlprefix  [0]{URL }%
\providecommand \Eprint [0]{\href }%
\providecommand \doibase [0]{http://dx.doi.org/}%
\providecommand \selectlanguage [0]{\@gobble}%
\providecommand \bibinfo  [0]{\@secondoftwo}%
\providecommand \bibfield  [0]{\@secondoftwo}%
\providecommand \translation [1]{[#1]}%
\providecommand \BibitemOpen [0]{}%
\providecommand \bibitemStop [0]{}%
\providecommand \bibitemNoStop [0]{.\EOS\space}%
\providecommand \EOS [0]{\spacefactor3000\relax}%
\providecommand \BibitemShut  [1]{\csname bibitem#1\endcsname}%
\let\auto@bib@innerbib\@empty
\bibitem [{\citenamefont {Basko}\ \emph {et~al.}(2006)\citenamefont {Basko},
  \citenamefont {Aleiner},\ and\ \citenamefont {Altschuler}}]{Basko06}%
  \BibitemOpen
  \bibfield  {author} {\bibinfo {author} {\bibfnamefont {D.}~\bibnamefont
  {Basko}}, \bibinfo {author} {\bibfnamefont {I.}~\bibnamefont {Aleiner}}, \
  and\ \bibinfo {author} {\bibfnamefont {B.}~\bibnamefont {Altschuler}},\
  }\href@noop {} {\bibfield  {journal} {\bibinfo  {journal} {Ann. Phys. (NY)}\
  }\textbf {\bibinfo {volume} {321}},\ \bibinfo {pages} {1126} (\bibinfo {year}
  {2006})}\BibitemShut {NoStop}%
\bibitem [{\citenamefont {Mehta}(1990)}]{Mehtabook}%
  \BibitemOpen
  \bibfield  {author} {\bibinfo {author} {\bibfnamefont {M.~L.}\ \bibnamefont
  {Mehta}},\ }\href@noop {} {\emph {\bibinfo {title} {Random Matrices}}}\
  (\bibinfo  {publisher} {Elsevier},\ \bibinfo {year} {1990})\BibitemShut
  {NoStop}%
\bibitem [{\citenamefont {Oganesyan}\ and\ \citenamefont
  {Huse}(2007)}]{Oganesyan07}%
  \BibitemOpen
  \bibfield  {author} {\bibinfo {author} {\bibfnamefont {V.}~\bibnamefont
  {Oganesyan}}\ and\ \bibinfo {author} {\bibfnamefont {D.~A.}\ \bibnamefont
  {Huse}},\ }\href {\doibase 10.1103/PhysRevB.75.155111} {\bibfield  {journal}
  {\bibinfo  {journal} {Phys. Rev. B}\ }\textbf {\bibinfo {volume} {75}},\
  \bibinfo {pages} {155111} (\bibinfo {year} {2007})}\BibitemShut {NoStop}%
\bibitem [{\citenamefont {Haake}(2010)}]{Haake}%
  \BibitemOpen
  \bibfield  {author} {\bibinfo {author} {\bibfnamefont {F.}~\bibnamefont
  {Haake}},\ }\href@noop {} {\emph {\bibinfo {title} {Quantum Signatures of
  Chaos}}}\ (\bibinfo  {publisher} {Springer, Berlin},\ \bibinfo {year}
  {2010})\BibitemShut {NoStop}%
\bibitem [{\citenamefont {G\'omez}\ \emph {et~al.}(2002)\citenamefont
  {G\'omez}, \citenamefont {Molina}, \citenamefont {Rela\~no},\ and\
  \citenamefont {Retamosa}}]{Gomez02}%
  \BibitemOpen
  \bibfield  {author} {\bibinfo {author} {\bibfnamefont {J.~M.~G.}\
  \bibnamefont {G\'omez}}, \bibinfo {author} {\bibfnamefont {R.~A.}\
  \bibnamefont {Molina}}, \bibinfo {author} {\bibfnamefont {A.}~\bibnamefont
  {Rela\~no}}, \ and\ \bibinfo {author} {\bibfnamefont {J.}~\bibnamefont
  {Retamosa}},\ }\href {\doibase 10.1103/PhysRevE.66.036209} {\bibfield
  {journal} {\bibinfo  {journal} {Phys. Rev. E}\ }\textbf {\bibinfo {volume}
  {66}},\ \bibinfo {pages} {036209} (\bibinfo {year} {2002})}\BibitemShut
  {NoStop}%
\bibitem [{\citenamefont {Bohigas}\ \emph {et~al.}(1984)\citenamefont
  {Bohigas}, \citenamefont {Giannoni},\ and\ \citenamefont
  {Schmit}}]{Bohigas84}%
  \BibitemOpen
  \bibfield  {author} {\bibinfo {author} {\bibfnamefont {O.}~\bibnamefont
  {Bohigas}}, \bibinfo {author} {\bibfnamefont {M.~J.}\ \bibnamefont
  {Giannoni}}, \ and\ \bibinfo {author} {\bibfnamefont {C.}~\bibnamefont
  {Schmit}},\ }\href {\doibase 10.1103/PhysRevLett.52.1} {\bibfield  {journal}
  {\bibinfo  {journal} {Phys. Rev. Lett.}\ }\textbf {\bibinfo {volume} {52}},\
  \bibinfo {pages} {1} (\bibinfo {year} {1984})}\BibitemShut {NoStop}%
\bibitem [{\citenamefont {Bohigas}\ \emph {et~al.}(1993)\citenamefont
  {Bohigas}, \citenamefont {Tomsovic},\ and\ \citenamefont
  {Ullmo}}]{Bohigas93}%
  \BibitemOpen
  \bibfield  {author} {\bibinfo {author} {\bibfnamefont {O.}~\bibnamefont
  {Bohigas}}, \bibinfo {author} {\bibfnamefont {S.}~\bibnamefont {Tomsovic}}, \
  and\ \bibinfo {author} {\bibfnamefont {D.}~\bibnamefont {Ullmo}},\ }\href
  {\doibase https://doi.org/10.1016/0370-1573(93)90109-Q} {\bibfield  {journal}
  {\bibinfo  {journal} {Physics Reports}\ }\textbf {\bibinfo {volume} {223}},\
  \bibinfo {pages} {43 } (\bibinfo {year} {1993})}\BibitemShut {NoStop}%
\bibitem [{\citenamefont {St\"ockmann}(1999)}]{Stockmann99}%
  \BibitemOpen
  \bibfield  {author} {\bibinfo {author} {\bibfnamefont {H.-J.}\ \bibnamefont
  {St\"ockmann}},\ }\href@noop {} {\emph {\bibinfo {title} {Quantum Chaos: An
  Introduction}}}\ (\bibinfo  {publisher} {University Press, Cambridge},\
  \bibinfo {year} {1999})\BibitemShut {NoStop}%
\bibitem [{\citenamefont {Guhr}\ \emph {et~al.}(1998)\citenamefont {Guhr},
  \citenamefont {M\"uller-Groeling},\ and\ \citenamefont
  {Weidenm\"uller}}]{Guhr98}%
  \BibitemOpen
  \bibfield  {author} {\bibinfo {author} {\bibfnamefont {T.}~\bibnamefont
  {Guhr}}, \bibinfo {author} {\bibfnamefont {A.}~\bibnamefont
  {M\"uller-Groeling}}, \ and\ \bibinfo {author} {\bibfnamefont {H.~A.}\
  \bibnamefont {Weidenm\"uller}},\ }\href {\doibase to be added} {\bibfield
  {journal} {\bibinfo  {journal} {Physics Reports}\ }\textbf {\bibinfo {volume}
  {299}},\ \bibinfo {pages} {189} (\bibinfo {year} {1998})}\BibitemShut
  {NoStop}%
\bibitem [{\citenamefont {Serbyn}\ and\ \citenamefont
  {Moore}(2016)}]{Serbyn16}%
  \BibitemOpen
  \bibfield  {author} {\bibinfo {author} {\bibfnamefont {M.}~\bibnamefont
  {Serbyn}}\ and\ \bibinfo {author} {\bibfnamefont {J.~E.}\ \bibnamefont
  {Moore}},\ }\href {\doibase 10.1103/PhysRevB.93.041424} {\bibfield  {journal}
  {\bibinfo  {journal} {Phys. Rev. B}\ }\textbf {\bibinfo {volume} {93}},\
  \bibinfo {pages} {041424} (\bibinfo {year} {2016})}\BibitemShut {NoStop}%
\bibitem [{\citenamefont {Bertrand}\ and\ \citenamefont
  {Garc\'{\i}a-Garc\'{\i}a}(2016)}]{Bertrand16}%
  \BibitemOpen
  \bibfield  {author} {\bibinfo {author} {\bibfnamefont {C.~L.}\ \bibnamefont
  {Bertrand}}\ and\ \bibinfo {author} {\bibfnamefont {A.~M.}\ \bibnamefont
  {Garc\'{\i}a-Garc\'{\i}a}},\ }\href {\doibase 10.1103/PhysRevB.94.144201}
  {\bibfield  {journal} {\bibinfo  {journal} {Phys. Rev. B}\ }\textbf {\bibinfo
  {volume} {94}},\ \bibinfo {pages} {144201} (\bibinfo {year}
  {2016})}\BibitemShut {NoStop}%
\bibitem [{\citenamefont {Sierant}\ and\ \citenamefont
  {Zakrzewski}(2019)}]{SierantPRB}%
  \BibitemOpen
  \bibfield  {author} {\bibinfo {author} {\bibfnamefont {P.}~\bibnamefont
  {Sierant}}\ and\ \bibinfo {author} {\bibfnamefont {J.}~\bibnamefont
  {Zakrzewski}},\ }\href {\doibase 10.1103/PhysRevB.99.104205} {\bibfield
  {journal} {\bibinfo  {journal} {Phys. Rev. B}\ }\textbf {\bibinfo {volume}
  {99}},\ \bibinfo {pages} {104205} (\bibinfo {year} {2019})}\BibitemShut
  {NoStop}%
\bibitem [{\citenamefont {Pechukas}(1983)}]{Pechukas83}%
  \BibitemOpen
  \bibfield  {author} {\bibinfo {author} {\bibfnamefont {P.}~\bibnamefont
  {Pechukas}},\ }\href {\doibase 10.1103/PhysRevLett.51.943} {\bibfield
  {journal} {\bibinfo  {journal} {Phys. Rev. Lett.}\ }\textbf {\bibinfo
  {volume} {51}},\ \bibinfo {pages} {943} (\bibinfo {year} {1983})}\BibitemShut
  {NoStop}%
\bibitem [{\citenamefont {Yukawa}(1985)}]{Yukawa85}%
  \BibitemOpen
  \bibfield  {author} {\bibinfo {author} {\bibfnamefont {T.}~\bibnamefont
  {Yukawa}},\ }\href {\doibase 10.1103/PhysRevLett.54.1883} {\bibfield
  {journal} {\bibinfo  {journal} {Phys. Rev. Lett.}\ }\textbf {\bibinfo
  {volume} {54}},\ \bibinfo {pages} {1883} (\bibinfo {year}
  {1985})}\BibitemShut {NoStop}%
\bibitem [{\citenamefont {Simons}\ and\ \citenamefont
  {Altshuler}(1993)}]{Simons93}%
  \BibitemOpen
  \bibfield  {author} {\bibinfo {author} {\bibfnamefont {B.~D.}\ \bibnamefont
  {Simons}}\ and\ \bibinfo {author} {\bibfnamefont {B.~L.}\ \bibnamefont
  {Altshuler}},\ }\href {\doibase 10.1103/PhysRevB.48.5422} {\bibfield
  {journal} {\bibinfo  {journal} {Phys. Rev. B}\ }\textbf {\bibinfo {volume}
  {48}},\ \bibinfo {pages} {5422} (\bibinfo {year} {1993})}\BibitemShut
  {NoStop}%
\bibitem [{\citenamefont {Zakrzewski}\ \emph {et~al.}(1993)\citenamefont
  {Zakrzewski}, \citenamefont {Delande},\ and\ \citenamefont
  {Ku\'s}}]{Zakrzewski93c}%
  \BibitemOpen
  \bibfield  {author} {\bibinfo {author} {\bibfnamefont {J.}~\bibnamefont
  {Zakrzewski}}, \bibinfo {author} {\bibfnamefont {D.}~\bibnamefont {Delande}},
  \ and\ \bibinfo {author} {\bibfnamefont {M.}~\bibnamefont {Ku\'s}},\ }\href
  {\doibase 10.1103/PhysRevE.47.1665} {\bibfield  {journal} {\bibinfo
  {journal} {Phys. Rev. E}\ }\textbf {\bibinfo {volume} {47}},\ \bibinfo
  {pages} {1665} (\bibinfo {year} {1993})}\BibitemShut {NoStop}%
\bibitem [{\citenamefont {Zakrzewski}\ and\ \citenamefont
  {Delande}(1993)}]{Zakrzewski93}%
  \BibitemOpen
  \bibfield  {author} {\bibinfo {author} {\bibfnamefont {J.}~\bibnamefont
  {Zakrzewski}}\ and\ \bibinfo {author} {\bibfnamefont {D.}~\bibnamefont
  {Delande}},\ }\href {\doibase 10.1103/PhysRevE.47.1650} {\bibfield  {journal}
  {\bibinfo  {journal} {Phys. Rev. E}\ }\textbf {\bibinfo {volume} {47}},\
  \bibinfo {pages} {1650} (\bibinfo {year} {1993})}\BibitemShut {NoStop}%
\bibitem [{\citenamefont {von Oppen}(1994)}]{vonOppen94}%
  \BibitemOpen
  \bibfield  {author} {\bibinfo {author} {\bibfnamefont {F.}~\bibnamefont {von
  Oppen}},\ }\href {\doibase 10.1103/PhysRevLett.73.798} {\bibfield  {journal}
  {\bibinfo  {journal} {Phys. Rev. Lett.}\ }\textbf {\bibinfo {volume} {73}},\
  \bibinfo {pages} {798} (\bibinfo {year} {1994})}\BibitemShut {NoStop}%
\bibitem [{\citenamefont {von Oppen}(1995)}]{vonOppen95}%
  \BibitemOpen
  \bibfield  {author} {\bibinfo {author} {\bibfnamefont {F.}~\bibnamefont {von
  Oppen}},\ }\href {\doibase 10.1103/PhysRevE.51.2647} {\bibfield  {journal}
  {\bibinfo  {journal} {Phys. Rev. E}\ }\textbf {\bibinfo {volume} {51}},\
  \bibinfo {pages} {2647} (\bibinfo {year} {1995})}\BibitemShut {NoStop}%
\bibitem [{\citenamefont {Fyodorov}\ and\ \citenamefont
  {Sommers}(1995)}]{Fyodorov95}%
  \BibitemOpen
  \bibfield  {author} {\bibinfo {author} {\bibfnamefont {Y.~V.}\ \bibnamefont
  {Fyodorov}}\ and\ \bibinfo {author} {\bibfnamefont {H.-J.}\ \bibnamefont
  {Sommers}},\ }\href {\doibase 10.1007/s002570050018} {\bibfield  {journal}
  {\bibinfo  {journal} {Zeitschrift f{\"u}r Physik B Condensed Matter}\
  }\textbf {\bibinfo {volume} {99}},\ \bibinfo {pages} {123} (\bibinfo {year}
  {1995})}\BibitemShut {NoStop}%
\bibitem [{\citenamefont {V~Fyodorov}(2011)}]{Fyodorov11}%
  \BibitemOpen
  \bibfield  {author} {\bibinfo {author} {\bibfnamefont {Y.}~\bibnamefont
  {V~Fyodorov}},\ }\href {\doibase 10.12693/APhysPolA.120.A-100} {\bibfield
  {journal} {\bibinfo  {journal} {Acta Physica Polonica A}\ }\textbf {\bibinfo
  {volume} {120}} (\bibinfo {year} {2011}),\
  10.12693/APhysPolA.120.A-100}\BibitemShut {NoStop}%
\bibitem [{\citenamefont {Filippone}\ \emph {et~al.}(2016)\citenamefont
  {Filippone}, \citenamefont {Brouwer}, \citenamefont {Eisert},\ and\
  \citenamefont {von Oppen}}]{Filippone16}%
  \BibitemOpen
  \bibfield  {author} {\bibinfo {author} {\bibfnamefont {M.}~\bibnamefont
  {Filippone}}, \bibinfo {author} {\bibfnamefont {P.~W.}\ \bibnamefont
  {Brouwer}}, \bibinfo {author} {\bibfnamefont {J.}~\bibnamefont {Eisert}}, \
  and\ \bibinfo {author} {\bibfnamefont {F.}~\bibnamefont {von Oppen}},\ }\href
  {\doibase 10.1103/PhysRevB.94.201112} {\bibfield  {journal} {\bibinfo
  {journal} {Phys. Rev. B}\ }\textbf {\bibinfo {volume} {94}},\ \bibinfo
  {pages} {201112} (\bibinfo {year} {2016})}\BibitemShut {NoStop}%
\bibitem [{\citenamefont {Monthus}(2017)}]{Monthus17}%
  \BibitemOpen
  \bibfield  {author} {\bibinfo {author} {\bibfnamefont {C.}~\bibnamefont
  {Monthus}},\ }\href {http://stacks.iop.org/1751-8121/50/i=9/a=095002}
  {\bibfield  {journal} {\bibinfo  {journal} {Journal of Physics A:
  Mathematical and Theoretical}\ }\textbf {\bibinfo {volume} {50}},\ \bibinfo
  {pages} {095002} (\bibinfo {year} {2017})}\BibitemShut {NoStop}%
\bibitem [{\citenamefont {Uhlmann}(1976)}]{Uhlmann76}%
  \BibitemOpen
  \bibfield  {author} {\bibinfo {author} {\bibfnamefont {A.}~\bibnamefont
  {Uhlmann}},\ }\href {\doibase https://doi.org/10.1016/0034-4877(76)90060-4}
  {\bibfield  {journal} {\bibinfo  {journal} {Reports on Mathematical Physics}\
  }\textbf {\bibinfo {volume} {9}},\ \bibinfo {pages} {273 } (\bibinfo {year}
  {1976})}\BibitemShut {NoStop}%
\bibitem [{\citenamefont {Zanardi}\ and\ \citenamefont
  {Paunkovi\ifmmode~\acute{c}\else \'{c}\fi{}}(2006)}]{Zanardi06}%
  \BibitemOpen
  \bibfield  {author} {\bibinfo {author} {\bibfnamefont {P.}~\bibnamefont
  {Zanardi}}\ and\ \bibinfo {author} {\bibfnamefont {N.}~\bibnamefont
  {Paunkovi\ifmmode~\acute{c}\else \'{c}\fi{}}},\ }\href {\doibase
  10.1103/PhysRevE.74.031123} {\bibfield  {journal} {\bibinfo  {journal} {Phys.
  Rev. E}\ }\textbf {\bibinfo {volume} {74}},\ \bibinfo {pages} {031123}
  (\bibinfo {year} {2006})}\BibitemShut {NoStop}%
\bibitem [{\citenamefont {Alhassid}\ and\ \citenamefont
  {Attias}(1995)}]{Alhassid95}%
  \BibitemOpen
  \bibfield  {author} {\bibinfo {author} {\bibfnamefont {Y.}~\bibnamefont
  {Alhassid}}\ and\ \bibinfo {author} {\bibfnamefont {H.}~\bibnamefont
  {Attias}},\ }\href {\doibase 10.1103/PhysRevLett.74.4635} {\bibfield
  {journal} {\bibinfo  {journal} {Phys. Rev. Lett.}\ }\textbf {\bibinfo
  {volume} {74}},\ \bibinfo {pages} {4635} (\bibinfo {year}
  {1995})}\BibitemShut {NoStop}%
\bibitem [{\citenamefont {H\"{u}bner}(1993)}]{Hubner93}%
  \BibitemOpen
  \bibfield  {author} {\bibinfo {author} {\bibfnamefont {M.}~\bibnamefont
  {H\"{u}bner}},\ }\href {\doibase
  http://dx.doi.org/10.1016/0375-9601(93)90668-P} {\bibfield  {journal}
  {\bibinfo  {journal} {Phys. Lett. A}\ }\textbf {\bibinfo {volume} {179}},\
  \bibinfo {pages} {226} (\bibinfo {year} {1993})}\BibitemShut {NoStop}%
\bibitem [{\citenamefont {Invernizzi}\ \emph {et~al.}(2008)\citenamefont
  {Invernizzi}, \citenamefont {Korbman}, \citenamefont {Campos~Venuti},\ and\
  \citenamefont {Paris}}]{Invernizzi08}%
  \BibitemOpen
  \bibfield  {author} {\bibinfo {author} {\bibfnamefont {C.}~\bibnamefont
  {Invernizzi}}, \bibinfo {author} {\bibfnamefont {M.}~\bibnamefont {Korbman}},
  \bibinfo {author} {\bibfnamefont {L.}~\bibnamefont {Campos~Venuti}}, \ and\
  \bibinfo {author} {\bibfnamefont {M.~G.~A.}\ \bibnamefont {Paris}},\ }\href
  {\doibase 10.1103/PhysRevA.78.042106} {\bibfield  {journal} {\bibinfo
  {journal} {Phys. Rev. A}\ }\textbf {\bibinfo {volume} {78}},\ \bibinfo
  {pages} {042106} (\bibinfo {year} {2008})}\BibitemShut {NoStop}%
\bibitem [{\citenamefont {Braun}\ \emph {et~al.}(2018)\citenamefont {Braun},
  \citenamefont {Adesso}, \citenamefont {Benatti}, \citenamefont {Floreanini},
  \citenamefont {Marzolino}, \citenamefont {Mitchell},\ and\ \citenamefont
  {Pirandola}}]{Braun18}%
  \BibitemOpen
  \bibfield  {author} {\bibinfo {author} {\bibfnamefont {D.}~\bibnamefont
  {Braun}}, \bibinfo {author} {\bibfnamefont {G.}~\bibnamefont {Adesso}},
  \bibinfo {author} {\bibfnamefont {F.}~\bibnamefont {Benatti}}, \bibinfo
  {author} {\bibfnamefont {R.}~\bibnamefont {Floreanini}}, \bibinfo {author}
  {\bibfnamefont {U.}~\bibnamefont {Marzolino}}, \bibinfo {author}
  {\bibfnamefont {M.~W.}\ \bibnamefont {Mitchell}}, \ and\ \bibinfo {author}
  {\bibfnamefont {S.}~\bibnamefont {Pirandola}},\ }\href {\doibase
  10.1103/RevModPhys.90.035006} {\bibfield  {journal} {\bibinfo  {journal}
  {Rev. Mod. Phys.}\ }\textbf {\bibinfo {volume} {90}},\ \bibinfo {pages}
  {035006} (\bibinfo {year} {2018})}\BibitemShut {NoStop}%
\bibitem [{\citenamefont {You}\ \emph {et~al.}(2007)\citenamefont {You},
  \citenamefont {Li},\ and\ \citenamefont {Gu}}]{You-Li-Gu}%
  \BibitemOpen
  \bibfield  {author} {\bibinfo {author} {\bibfnamefont {W.-L.}\ \bibnamefont
  {You}}, \bibinfo {author} {\bibfnamefont {Y.-W.}\ \bibnamefont {Li}}, \ and\
  \bibinfo {author} {\bibfnamefont {S.-J.}\ \bibnamefont {Gu}},\ }\href
  {\doibase 10.1103/PhysRevE.76.022101} {\bibfield  {journal} {\bibinfo
  {journal} {Phys. Rev. E}\ }\textbf {\bibinfo {volume} {76}},\ \bibinfo
  {pages} {022101} (\bibinfo {year} {2007})}\BibitemShut {NoStop}%
\bibitem [{\citenamefont {Zanardi}\ \emph {et~al.}(2008)\citenamefont
  {Zanardi}, \citenamefont {Paris},\ and\ \citenamefont
  {Campos~Venuti}}]{Paris}%
  \BibitemOpen
  \bibfield  {author} {\bibinfo {author} {\bibfnamefont {P.}~\bibnamefont
  {Zanardi}}, \bibinfo {author} {\bibfnamefont {M.~G.~A.}\ \bibnamefont
  {Paris}}, \ and\ \bibinfo {author} {\bibfnamefont {L.}~\bibnamefont
  {Campos~Venuti}},\ }\href {\doibase 10.1103/PhysRevA.78.042105} {\bibfield
  {journal} {\bibinfo  {journal} {Phys. Rev. A}\ }\textbf {\bibinfo {volume}
  {78}},\ \bibinfo {pages} {042105} (\bibinfo {year} {2008})}\BibitemShut
  {NoStop}%
\bibitem [{\citenamefont {Salvatori}\ \emph {et~al.}(2014)\citenamefont
  {Salvatori}, \citenamefont {Mandarino},\ and\ \citenamefont
  {Paris}}]{Salvatori2014}%
  \BibitemOpen
  \bibfield  {author} {\bibinfo {author} {\bibfnamefont {G.}~\bibnamefont
  {Salvatori}}, \bibinfo {author} {\bibfnamefont {A.}~\bibnamefont
  {Mandarino}}, \ and\ \bibinfo {author} {\bibfnamefont {M.~G.~A.}\
  \bibnamefont {Paris}},\ }\href {\doibase 10.1103/PhysRevA.90.022111}
  {\bibfield  {journal} {\bibinfo  {journal} {Phys. Rev. A}\ }\textbf {\bibinfo
  {volume} {90}},\ \bibinfo {pages} {022111} (\bibinfo {year}
  {2014})}\BibitemShut {NoStop}%
\bibitem [{\citenamefont {Bina}\ \emph {et~al.}(2016)\citenamefont {Bina},
  \citenamefont {Amelio},\ and\ \citenamefont {Paris}}]{Bina2016}%
  \BibitemOpen
  \bibfield  {author} {\bibinfo {author} {\bibfnamefont {M.}~\bibnamefont
  {Bina}}, \bibinfo {author} {\bibfnamefont {I.}~\bibnamefont {Amelio}}, \ and\
  \bibinfo {author} {\bibfnamefont {M.~G.~A.}\ \bibnamefont {Paris}},\ }\href
  {\doibase 10.1103/PhysRevE.93.052118} {\bibfield  {journal} {\bibinfo
  {journal} {Phys. Rev. E}\ }\textbf {\bibinfo {volume} {93}},\ \bibinfo
  {pages} {052118} (\bibinfo {year} {2016})}\BibitemShut {NoStop}%
\bibitem [{\citenamefont {Boyajian}\ \emph {et~al.}(2016)\citenamefont
  {Boyajian}, \citenamefont {Skotiniotis}, \citenamefont {D\"ur},\ and\
  \citenamefont {Kraus}}]{Kraus16}%
  \BibitemOpen
  \bibfield  {author} {\bibinfo {author} {\bibfnamefont {W.~L.}\ \bibnamefont
  {Boyajian}}, \bibinfo {author} {\bibfnamefont {M.}~\bibnamefont
  {Skotiniotis}}, \bibinfo {author} {\bibfnamefont {W.}~\bibnamefont {D\"ur}},
  \ and\ \bibinfo {author} {\bibfnamefont {B.}~\bibnamefont {Kraus}},\ }\href
  {\doibase 10.1103/PhysRevA.94.062326} {\bibfield  {journal} {\bibinfo
  {journal} {Phys. Rev. A}\ }\textbf {\bibinfo {volume} {94}},\ \bibinfo
  {pages} {062326} (\bibinfo {year} {2016})}\BibitemShut {NoStop}%
\bibitem [{\citenamefont {Mehboudi}\ \emph {et~al.}(2016)\citenamefont
  {Mehboudi}, \citenamefont {Correa},\ and\ \citenamefont
  {Sanpera}}]{sanpera_2016}%
  \BibitemOpen
  \bibfield  {author} {\bibinfo {author} {\bibfnamefont {M.}~\bibnamefont
  {Mehboudi}}, \bibinfo {author} {\bibfnamefont {L.~A.}\ \bibnamefont
  {Correa}}, \ and\ \bibinfo {author} {\bibfnamefont {A.}~\bibnamefont
  {Sanpera}},\ }\href {\doibase 10.1103/PhysRevA.94.042121} {\bibfield
  {journal} {\bibinfo  {journal} {Phys. Rev. A}\ }\textbf {\bibinfo {volume}
  {94}},\ \bibinfo {pages} {042121} (\bibinfo {year} {2016})}\BibitemShut
  {NoStop}%
\bibitem [{\citenamefont {Zanardi}\ \emph {et~al.}(2007)\citenamefont
  {Zanardi}, \citenamefont {Quan}, \citenamefont {Wang},\ and\ \citenamefont
  {Sun}}]{Zanardi07}%
  \BibitemOpen
  \bibfield  {author} {\bibinfo {author} {\bibfnamefont {P.}~\bibnamefont
  {Zanardi}}, \bibinfo {author} {\bibfnamefont {H.~T.}\ \bibnamefont {Quan}},
  \bibinfo {author} {\bibfnamefont {X.}~\bibnamefont {Wang}}, \ and\ \bibinfo
  {author} {\bibfnamefont {C.~P.}\ \bibnamefont {Sun}},\ }\href {\doibase
  10.1103/PhysRevA.75.032109} {\bibfield  {journal} {\bibinfo  {journal} {Phys.
  Rev. A}\ }\textbf {\bibinfo {volume} {75}},\ \bibinfo {pages} {032109}
  (\bibinfo {year} {2007})}\BibitemShut {NoStop}%
\bibitem [{\citenamefont {Quan}\ and\ \citenamefont
  {Cucchietti}(2009)}]{Quan09}%
  \BibitemOpen
  \bibfield  {author} {\bibinfo {author} {\bibfnamefont {H.~T.}\ \bibnamefont
  {Quan}}\ and\ \bibinfo {author} {\bibfnamefont {F.~M.}\ \bibnamefont
  {Cucchietti}},\ }\href {\doibase 10.1103/PhysRevE.79.031101} {\bibfield
  {journal} {\bibinfo  {journal} {Phys. Rev. E}\ }\textbf {\bibinfo {volume}
  {79}},\ \bibinfo {pages} {031101} (\bibinfo {year} {2009})}\BibitemShut
  {NoStop}%
\bibitem [{\citenamefont {Sirker}(2010)}]{Sirker10}%
  \BibitemOpen
  \bibfield  {author} {\bibinfo {author} {\bibfnamefont {J.}~\bibnamefont
  {Sirker}},\ }\href {\doibase 10.1103/PhysRevLett.105.117203} {\bibfield
  {journal} {\bibinfo  {journal} {Phys. Rev. Lett.}\ }\textbf {\bibinfo
  {volume} {105}},\ \bibinfo {pages} {117203} (\bibinfo {year}
  {2010})}\BibitemShut {NoStop}%
\bibitem [{\citenamefont {Rams}\ \emph {et~al.}(2018)\citenamefont {Rams},
  \citenamefont {Sierant}, \citenamefont {Dutta}, \citenamefont {Horodecki},\
  and\ \citenamefont {Zakrzewski}}]{Rams18}%
  \BibitemOpen
  \bibfield  {author} {\bibinfo {author} {\bibfnamefont {M.~M.}\ \bibnamefont
  {Rams}}, \bibinfo {author} {\bibfnamefont {P.}~\bibnamefont {Sierant}},
  \bibinfo {author} {\bibfnamefont {O.}~\bibnamefont {Dutta}}, \bibinfo
  {author} {\bibfnamefont {P.}~\bibnamefont {Horodecki}}, \ and\ \bibinfo
  {author} {\bibfnamefont {J.}~\bibnamefont {Zakrzewski}},\ }\href {\doibase
  10.1103/PhysRevX.8.021022} {\bibfield  {journal} {\bibinfo  {journal} {Phys.
  Rev. X}\ }\textbf {\bibinfo {volume} {8}},\ \bibinfo {pages} {021022}
  (\bibinfo {year} {2018})}\BibitemShut {NoStop}%
\bibitem [{\citenamefont {Hu}\ \emph {et~al.}(2016)\citenamefont {Hu},
  \citenamefont {Xue}, \citenamefont {Li}, \citenamefont {Zhang},\ and\
  \citenamefont {Ren}}]{Hu16}%
  \BibitemOpen
  \bibfield  {author} {\bibinfo {author} {\bibfnamefont {T.}~\bibnamefont
  {Hu}}, \bibinfo {author} {\bibfnamefont {K.}~\bibnamefont {Xue}}, \bibinfo
  {author} {\bibfnamefont {X.}~\bibnamefont {Li}}, \bibinfo {author}
  {\bibfnamefont {Y.}~\bibnamefont {Zhang}}, \ and\ \bibinfo {author}
  {\bibfnamefont {H.}~\bibnamefont {Ren}},\ }\href {\doibase
  10.1103/PhysRevE.94.052119} {\bibfield  {journal} {\bibinfo  {journal} {Phys.
  Rev. E}\ }\textbf {\bibinfo {volume} {94}},\ \bibinfo {pages} {052119}
  (\bibinfo {year} {2016})}\BibitemShut {NoStop}%
\bibitem [{\citenamefont {{Maksymov}}\ \emph {et~al.}(2019)\citenamefont
  {{Maksymov}}, \citenamefont {{Sierant}},\ and\ \citenamefont
  {{Zakrzewski}}}]{Maksymov19}%
  \BibitemOpen
  \bibfield  {author} {\bibinfo {author} {\bibfnamefont {A.}~\bibnamefont
  {{Maksymov}}}, \bibinfo {author} {\bibfnamefont {P.}~\bibnamefont
  {{Sierant}}}, \ and\ \bibinfo {author} {\bibfnamefont {J.}~\bibnamefont
  {{Zakrzewski}}},\ }\href@noop {} {\bibfield  {journal} {\bibinfo  {journal}
  {arXiv e-prints}\ ,\ \bibinfo {eid} {arXiv:1904.05057}} (\bibinfo {year}
  {2019})},\ \Eprint {http://arxiv.org/abs/1904.05057} {arXiv:1904.05057
  [cond-mat.dis-nn]} \BibitemShut {NoStop}%
\bibitem [{Sup()}]{Suppl}%
  \BibitemOpen
  \href@noop {} {}\bibinfo {note} {See Supplemental Material at [URL will be
  inserted by publisher] for details on the derivation of fidelity
  susceptibility distributions both for GOE and GUE.}\BibitemShut {Stop}%
\bibitem [{\citenamefont {Poli}\ \emph {et~al.}(2009)\citenamefont {Poli},
  \citenamefont {Savin}, \citenamefont {Legrand},\ and\ \citenamefont
  {Mortessagne}}]{Poli09}%
  \BibitemOpen
  \bibfield  {author} {\bibinfo {author} {\bibfnamefont {C.}~\bibnamefont
  {Poli}}, \bibinfo {author} {\bibfnamefont {D.~V.}\ \bibnamefont {Savin}},
  \bibinfo {author} {\bibfnamefont {O.}~\bibnamefont {Legrand}}, \ and\
  \bibinfo {author} {\bibfnamefont {F.}~\bibnamefont {Mortessagne}},\ }\href
  {\doibase 10.1103/PhysRevE.80.046203} {\bibfield  {journal} {\bibinfo
  {journal} {Phys. Rev. E}\ }\textbf {\bibinfo {volume} {80}},\ \bibinfo
  {pages} {046203} (\bibinfo {year} {2009})}\BibitemShut {NoStop}%
\bibitem [{\citenamefont {Fyodorov}\ and\ \citenamefont
  {Savin}(2012)}]{Fyodorov12}%
  \BibitemOpen
  \bibfield  {author} {\bibinfo {author} {\bibfnamefont {Y.~V.}\ \bibnamefont
  {Fyodorov}}\ and\ \bibinfo {author} {\bibfnamefont {D.~V.}\ \bibnamefont
  {Savin}},\ }\href {\doibase 10.1103/PhysRevLett.108.184101} {\bibfield
  {journal} {\bibinfo  {journal} {Phys. Rev. Lett.}\ }\textbf {\bibinfo
  {volume} {108}},\ \bibinfo {pages} {184101} (\bibinfo {year}
  {2012})}\BibitemShut {NoStop}%
\bibitem [{\citenamefont {Fyodorov}\ and\ \citenamefont
  {Nock}(2015)}]{Fyodorov15}%
  \BibitemOpen
  \bibfield  {author} {\bibinfo {author} {\bibfnamefont {Y.~V.}\ \bibnamefont
  {Fyodorov}}\ and\ \bibinfo {author} {\bibfnamefont {A.}~\bibnamefont
  {Nock}},\ }\href {\doibase 10.1007/s10955-015-1209-x} {\bibfield  {journal}
  {\bibinfo  {journal} {Journal of Statistical Physics}\ }\textbf {\bibinfo
  {volume} {159}},\ \bibinfo {pages} {731} (\bibinfo {year}
  {2015})}\BibitemShut {NoStop}%
\bibitem [{\citenamefont {Delannay}\ and\ \citenamefont
  {Le~Ca\"er}(2000)}]{Delannay00}%
  \BibitemOpen
  \bibfield  {author} {\bibinfo {author} {\bibfnamefont {R.}~\bibnamefont
  {Delannay}}\ and\ \bibinfo {author} {\bibfnamefont {G.}~\bibnamefont
  {Le~Ca\"er}},\ }\href {\doibase 10.1103/PhysRevE.62.1526} {\bibfield
  {journal} {\bibinfo  {journal} {Phys. Rev. E}\ }\textbf {\bibinfo {volume}
  {62}},\ \bibinfo {pages} {1526} (\bibinfo {year} {2000})}\BibitemShut
  {NoStop}%
\bibitem [{\citenamefont {Br{\'e}zin}\ and\ \citenamefont
  {Hikami}(2000)}]{Brezin00}%
  \BibitemOpen
  \bibfield  {author} {\bibinfo {author} {\bibfnamefont {E.}~\bibnamefont
  {Br{\'e}zin}}\ and\ \bibinfo {author} {\bibfnamefont {S.}~\bibnamefont
  {Hikami}},\ }\href {\doibase 10.1007/s002200000256} {\bibfield  {journal}
  {\bibinfo  {journal} {Communications in Mathematical Physics}\ }\textbf
  {\bibinfo {volume} {214}},\ \bibinfo {pages} {111} (\bibinfo {year}
  {2000})}\BibitemShut {NoStop}%
\bibitem [{\citenamefont {Fyodorov}\ and\ \citenamefont
  {Strahov}(2003)}]{Fyodorov03}%
  \BibitemOpen
  \bibfield  {author} {\bibinfo {author} {\bibfnamefont {Y.~V.}\ \bibnamefont
  {Fyodorov}}\ and\ \bibinfo {author} {\bibfnamefont {E.}~\bibnamefont
  {Strahov}},\ }\href {http://stacks.iop.org/0305-4470/36/i=12/a=320}
  {\bibfield  {journal} {\bibinfo  {journal} {Journal of Physics A:
  Mathematical and General}\ }\textbf {\bibinfo {volume} {36}},\ \bibinfo
  {pages} {3203} (\bibinfo {year} {2003})}\BibitemShut {NoStop}%
\bibitem [{\citenamefont {Strahov}\ and\ \citenamefont
  {Fyodorov}(2003)}]{Strahov03}%
  \BibitemOpen
  \bibfield  {author} {\bibinfo {author} {\bibfnamefont {E.}~\bibnamefont
  {Strahov}}\ and\ \bibinfo {author} {\bibfnamefont {Y.~V.}\ \bibnamefont
  {Fyodorov}},\ }\href {\doibase 10.1007/s00220-003-0938-x} {\bibfield
  {journal} {\bibinfo  {journal} {Communications in Mathematical Physics}\
  }\textbf {\bibinfo {volume} {241}},\ \bibinfo {pages} {343} (\bibinfo {year}
  {2003})}\BibitemShut {NoStop}%
\bibitem [{\citenamefont {Mehta}\ and\ \citenamefont
  {Normand}(1998)}]{Mehta98}%
  \BibitemOpen
  \bibfield  {author} {\bibinfo {author} {\bibfnamefont {M.~L.}\ \bibnamefont
  {Mehta}}\ and\ \bibinfo {author} {\bibfnamefont {J.-M.}\ \bibnamefont
  {Normand}},\ }\href {http://stacks.iop.org/0305-4470/31/i=23/a=018}
  {\bibfield  {journal} {\bibinfo  {journal} {Journal of Physics A:
  Mathematical and General}\ }\textbf {\bibinfo {volume} {31}},\ \bibinfo
  {pages} {5377} (\bibinfo {year} {1998})}\BibitemShut {NoStop}%
\bibitem [{\citenamefont {Cicuta}\ and\ \citenamefont
  {Mehta}(2000)}]{Cicuta00}%
  \BibitemOpen
  \bibfield  {author} {\bibinfo {author} {\bibfnamefont {G.~M.}\ \bibnamefont
  {Cicuta}}\ and\ \bibinfo {author} {\bibfnamefont {M.~L.}\ \bibnamefont
  {Mehta}},\ }\href {http://stacks.iop.org/0305-4470/33/i=45/a=302} {\bibfield
  {journal} {\bibinfo  {journal} {Journal of Physics A: Mathematical and
  General}\ }\textbf {\bibinfo {volume} {33}},\ \bibinfo {pages} {8029}
  (\bibinfo {year} {2000})}\BibitemShut {NoStop}%
\bibitem [{\citenamefont {Gaspard}\ \emph {et~al.}(1990)\citenamefont
  {Gaspard}, \citenamefont {Rice}, \citenamefont {Mikeska},\ and\ \citenamefont
  {Nakamura}}]{Gaspard90}%
  \BibitemOpen
  \bibfield  {author} {\bibinfo {author} {\bibfnamefont {P.}~\bibnamefont
  {Gaspard}}, \bibinfo {author} {\bibfnamefont {S.~A.}\ \bibnamefont {Rice}},
  \bibinfo {author} {\bibfnamefont {H.~J.}\ \bibnamefont {Mikeska}}, \ and\
  \bibinfo {author} {\bibfnamefont {K.}~\bibnamefont {Nakamura}},\ }\href
  {\doibase 10.1103/PhysRevA.42.4015} {\bibfield  {journal} {\bibinfo
  {journal} {Phys. Rev. A}\ }\textbf {\bibinfo {volume} {42}},\ \bibinfo
  {pages} {4015} (\bibinfo {year} {1990})}\BibitemShut {NoStop}%
\bibitem [{\citenamefont {Gu}\ and\ \citenamefont {Yu}(2014)}]{Shi-Jian14}%
  \BibitemOpen
  \bibfield  {author} {\bibinfo {author} {\bibfnamefont {S.-J.}\ \bibnamefont
  {Gu}}\ and\ \bibinfo {author} {\bibfnamefont {W.~C.}\ \bibnamefont {Yu}},\
  }\href {http://stacks.iop.org/0295-5075/108/i=2/a=20002} {\bibfield
  {journal} {\bibinfo  {journal} {EPL (Europhysics Letters)}\ }\textbf
  {\bibinfo {volume} {108}},\ \bibinfo {pages} {20002} (\bibinfo {year}
  {2014})}\BibitemShut {NoStop}%
\bibitem [{\citenamefont {Ernst}\ \emph {et~al.}(2010)\citenamefont {Ernst},
  \citenamefont {Gotze}, \citenamefont {Krauser}, \citenamefont {Pyka},
  \citenamefont {Luhmann}, \citenamefont {Pfannkuche},\ and\ \citenamefont
  {Sengstock}}]{Ernst10}%
  \BibitemOpen
  \bibfield  {author} {\bibinfo {author} {\bibfnamefont {P.~T.}\ \bibnamefont
  {Ernst}}, \bibinfo {author} {\bibfnamefont {S.}~\bibnamefont {Gotze}},
  \bibinfo {author} {\bibfnamefont {J.~S.}\ \bibnamefont {Krauser}}, \bibinfo
  {author} {\bibfnamefont {K.}~\bibnamefont {Pyka}}, \bibinfo {author}
  {\bibfnamefont {D.-S.}\ \bibnamefont {Luhmann}}, \bibinfo {author}
  {\bibfnamefont {D.}~\bibnamefont {Pfannkuche}}, \ and\ \bibinfo {author}
  {\bibfnamefont {K.}~\bibnamefont {Sengstock}},\ }\href {\doibase
  10.1038/nphys1476} {\bibfield  {journal} {\bibinfo  {journal} {Nat Phys}\
  }\textbf {\bibinfo {volume} {6}},\ \bibinfo {pages} {56} (\bibinfo {year}
  {2010})}\BibitemShut {NoStop}%
\bibitem [{\citenamefont {Cl\'ement}\ \emph {et~al.}(2010)\citenamefont
  {Cl\'ement}, \citenamefont {Fabbri}, \citenamefont {Fallani}, \citenamefont
  {Fort},\ and\ \citenamefont {Inguscio}}]{Inguscio}%
  \BibitemOpen
  \bibfield  {author} {\bibinfo {author} {\bibfnamefont {D.}~\bibnamefont
  {Cl\'ement}}, \bibinfo {author} {\bibfnamefont {N.}~\bibnamefont {Fabbri}},
  \bibinfo {author} {\bibfnamefont {L.}~\bibnamefont {Fallani}}, \bibinfo
  {author} {\bibfnamefont {C.}~\bibnamefont {Fort}}, \ and\ \bibinfo {author}
  {\bibfnamefont {M.}~\bibnamefont {Inguscio}},\ }\href {\doibase
  10.1007/s10909-009-0040-7} {\bibfield  {journal} {\bibinfo  {journal} {J. Low
  Temp. Phys.}\ }\textbf {\bibinfo {volume} {158}},\ \bibinfo {pages} {5}
  (\bibinfo {year} {2010})}\BibitemShut {NoStop}%
\bibitem [{\citenamefont {Zhang}\ \emph {et~al.}(2008)\citenamefont {Zhang},
  \citenamefont {Peng}, \citenamefont {Rajendran},\ and\ \citenamefont
  {Suter}}]{Zhang2008}%
  \BibitemOpen
  \bibfield  {author} {\bibinfo {author} {\bibfnamefont {J.}~\bibnamefont
  {Zhang}}, \bibinfo {author} {\bibfnamefont {X.}~\bibnamefont {Peng}},
  \bibinfo {author} {\bibfnamefont {N.}~\bibnamefont {Rajendran}}, \ and\
  \bibinfo {author} {\bibfnamefont {D.}~\bibnamefont {Suter}},\ }\href
  {\doibase 10.1103/PhysRevLett.100.100501} {\bibfield  {journal} {\bibinfo
  {journal} {Phys. Rev. Lett.}\ }\textbf {\bibinfo {volume} {100}},\ \bibinfo
  {pages} {100501} (\bibinfo {year} {2008})}\BibitemShut {NoStop}%
\bibitem [{\citenamefont {Islam}\ \emph {et~al.}(2015)\citenamefont {Islam},
  \citenamefont {Ma}, \citenamefont {Preiss}, \citenamefont {Eric~Tai},
  \citenamefont {Lukin}, \citenamefont {Rispoli},\ and\ \citenamefont
  {Greiner}}]{Islam15}%
  \BibitemOpen
  \bibfield  {author} {\bibinfo {author} {\bibfnamefont {R.}~\bibnamefont
  {Islam}}, \bibinfo {author} {\bibfnamefont {R.}~\bibnamefont {Ma}}, \bibinfo
  {author} {\bibfnamefont {P.~M.}\ \bibnamefont {Preiss}}, \bibinfo {author}
  {\bibfnamefont {M.}~\bibnamefont {Eric~Tai}}, \bibinfo {author}
  {\bibfnamefont {A.}~\bibnamefont {Lukin}}, \bibinfo {author} {\bibfnamefont
  {M.}~\bibnamefont {Rispoli}}, \ and\ \bibinfo {author} {\bibfnamefont
  {M.}~\bibnamefont {Greiner}},\ }\href {https://doi.org/10.1038/nature15750}
  {\bibfield  {journal} {\bibinfo  {journal} {Nature}\ }\textbf {\bibinfo
  {volume} {528}},\ \bibinfo {pages} {77 EP } (\bibinfo {year}
  {2015})}\BibitemShut {NoStop}%
\end{thebibliography}
%


\newcommand{\snum}{S}

\renewcommand{\theequation}{\snum.\arabic{equation}}

\setcounter{equation}{0}

\begin{widetext}

 \section{Supplementary material to \\ ``Fidelity susceptibility in Gaussian Random Ensembles''}
 
 \subsection{Derivation of formulas \eqref{s6} and \eqref{s6a}}
To obtain the equation \eqref{s6} we 
use Fourier representation for $\delta(\chi-\chi_n)$ rewriting \eqref{probability} as
\begin{equation}\label{s1}
P(\chi,E)=\frac{1}{2\pi N\rho(E)}\sum_{n=1}^N\int_{-\infty}^{\infty}d\omega 
e^{-i\omega\chi}\left\langle\delta(E-E_n)\exp\left(i\omega\sum_{m\ne n}
\frac{|H_{1,nm}|^2}{(E_n-E_m)^2} \right)  \right\rangle. 
\end{equation}
The averaging over $H_1$ with the probability density (\ref{pH1}) reduces to a Gaussian integral and gives,
\begin{equation}\label{s2}
P(\chi,E)=\frac{1}{2\pi N\rho(E)}\sum_{n=1}^N
\int_{-\infty}^{\infty}d\omega e^{-i\omega\chi}\left\langle\delta(E-E_n)
\prod_{m\ne n}\left(1-\frac{2i\omega J^2}{\beta(E_n-E_m)^2} \right)^{-\frac{\beta}{2}} \right\rangle. 
\end{equation}
The remaining averaging over 
the distribution $P(H_0)$ reduces to average over 
eigenvalues $E_1,\ldots, E_N$ of $H_0$
\begin{eqnarray}\label{s3}
P(\chi,E)\sim\sum_{n=1}^N\int_{-\infty}^{\infty}
d\omega e^{-i\omega\chi} \int \prod_{j=1}^N dE_j \delta(E-E_n) 
 \prod_{k<l}
\left| E_k-E_l\right|^\beta   \mathrm{e}^{-\frac{\beta}{4J^2}\sum_k E_k^2}
\prod_{m\ne n}\left(1-\frac{2i\omega J^2}{\beta(E_n-E_m)^2} \right)^{-\frac{\beta}{2}}.
\end{eqnarray}
Now we can perform the integral over $E_n$. There are $N$ such integrals due to the summation from $n=1$ to $n=N$ at the beginning of the formula. So let's take $E_n=E_1$. Due to the delta function we can substitute $E_1=E$ and rewrite the averaging over the eigenvalues as 
\begin{eqnarray}\label{s4}
&&\int dE_1\cdots dE_N\delta(E-E_1)\prod_{k<l}\left| E_k-E_l\right|^\beta\exp\left(-\frac{\beta}{4J^2}\sum_k E_k^2\right)\prod_{m\ne n}\left(1-\frac{2i\omega J^2}{\beta(E_n-E_m)^2} \right)^{-\frac{\beta}{2}} = \nonumber \\
&=& e^{-\frac{\beta E^2}{4J^2}}\int dE_2\cdots dE_N\prod_{m=2}\left| E-E_m\right|^\beta\prod_{m=2}\left(1-\frac{2i\omega J^2}{\beta(E-E_m)^2} \right)^{-\frac{\beta}{2}}\prod_{2\le k< l}\left| E_k-E_l\right|^\beta\exp\left(-\frac{\beta}{4J^2}\sum_{k=2}E_k^2\right)= 
\nonumber \\
&=&e^{-\frac{\beta E^2}{4J^2}}\left\langle\prod_{m=2}\left(1-\frac{2i\omega J^2}{\beta(E-E_m)^2} \right)^{-\frac{\beta}{2}}\left| E-E_m\right|^\beta \right\rangle, 
\end{eqnarray}
where the averaging goes over the joint probability of the remaining eigenvalues $E_2,\ldots,E_n$.    

\noindent At the center of the spectrum $E=0$ the averaged quantity reads
\begin{equation}\label{s5}
\prod_{m=2}\left[ \frac{\left|E_m\right|}{\left(1-\frac{2i\omega J^2}{\beta E_m^2}
\right)^{\frac{1}{2}}}\right]^\beta=
\left[\frac{\det \bar{H}^2}{\det\left(\bar{H}^2-\frac{2i\omega 
J^2}{\beta}\right)^{\frac{1}{2}}} \right]^\beta 
\end{equation}
Plugging \eqref{s5} into \eqref{s4},  we finally arrive at \eqref{s6}.

The denominator in \eqref{s6} can be expressed in the form of a Gaussian integral
\begin{equation}\label{s7}
\det\left(\bar{H}^2-\frac{2i\omega J^2}{\beta}\right)^{-\frac{\beta}{2}}\sim\int d\mathbf{z}\, \exp\left[-\mathbf{z}^\dagger\left(\bar{H}^2-\frac{2i\omega J^2}{\beta}\right)\mathbf{ z} \right]=\int d\mathbf{z}\, \exp\left(-\mathbf{z}^\dagger\bar{H}^2\mathbf{z}\right)e^{\frac{2i\omega J^2|\mathbf{ z}|^2}{\beta}}, 
\end{equation}
where $\mathbf{ z}$ is a $N-1$-dimensional vector, real for $\beta=1$ and complex for $\beta=2$.
Due to the invariance of the ensembles with respect to appropriate ($O(N-1)$ or $U(N-1)$) 
rotations the average does not depend on the direction of $\mathbf{z}$, but only on 
its norm $|\mathbf{z}|^2$ 
\begin{equation}\label{s9}
P(\chi)\sim\int_{-\infty}^{\infty}d\omega e^{-i\omega\chi}\left\langle\det \bar{H}^2\int d\mathbf{z}\, \exp\left(-|\mathbf{z}|^2 X\right)e^{\frac{2i\omega J^2|\mathbf{ z}|^2}{\beta}}\right\rangle, 
\end{equation} 
where $X$ is some quadratic form in the elements of 
$\bar{H}$ specified below. In the spherical coordinates
$d\mathbf{z}\sim drr^{\beta(N-1)\beta-1}$ (where $r:=|\mathbf{z}|$), 
integrating over $\omega$ results in $\delta\left(\chi-2J^2r^2/\beta\right)$
and thus we arrive at \eqref{s6a}.

\subsection{Fidelity susceptibility distribution for GOE}
For GOE ($\beta=1$), choosing $\mathbf{z} = r[1,0,0..]^T$ we rewrite the average in  \eqref{s6a} as
\begin{equation}
 \left \langle \mathrm{det}\bar H^{2} \mathrm{e}^{-r^2 X}
 \right \rangle_{N-1} =
 \int d\bar H_{11}\mathrm{e}^{-A \bar H_{11}^2}\prod_{j=2}^{N-1} 
 d \bar H_{1j}\mathrm{e}^{-B \sum_{j=2}^{N-1} \bar H_{1j}^2}
  \mathrm{det}\bar H^{2} D^{N-2}V,
  \label{eq: 2}
\end{equation}
with $A = \frac{1}{4J^2} +r^2$, $B=\frac{1}{2J^2} +r^2$, $X=\sum_{j=1}^{N-1} | \bar H_{1j}|^2$ 
and \begin{equation}\label{s13}
\bar{H}=\left[
\begin{array}{cc}
H_{11} & H_{1j} \\ 
H_{1k} & V
\end{array} 
\right].
\end{equation}
The block $V$ is itself a $(N-2)\times(N-2)$ GOE matrix (with the GOE density 
$\mathrm{D}^{N-2}V=\prod_{k<j} dV_{kj}\exp(-\frac{1}{4J^2}\Tr V^2)$).
Using the general formula for the determinant of a block matrix
\begin{equation}
\det\left[\begin{array}{cc}
\mathbf{A} & \mathbf{B} \\ 
\mathbf{C} & \mathbf{D}
\end{array} \right]=\det\left(\mathbf{A}-\mathbf{B}\mathbf{D}^{-1}\mathbf{C}\right)\det\mathbf{D}
\end{equation}
we get (since the upper diagonal block is in fact one-dimensional, $\mathbf{A}=H_{11}$),
\begin{equation}\label{s15}
\det\bar{H}=\det V\left(\bar{H}_{11}-\sum_{j,k=2}^{N-1}\bar{H}_{1j}V^{-1}_{jk}\bar{H}_{1k}\right).
\end{equation}
Thus, \eqref{eq: 2} becomes
\begin{equation}
 \left \langle \mathrm{det}\bar H^{2} \mathrm{e}^{-r^2 X}
 \right \rangle_{N-1}=
 \int d\bar H_{11} \mathrm{e}^{-A \bar H_{11}^2}
 \prod_{j=2}^{N-1} d \bar H_{1j} \mathrm{e}^{-B \sum_{j=2}^{N-1} | \bar H_{1j}|^2}
   \left \langle \mathrm{det}V^{2} \left( \bar H_{11}- 
   \sum_{j,k=2}^{N-1} \bar H_{1j} V^{-1}_{jk} \bar H_{1k} \right)^{2} \right \rangle_{N-2},
   \label{eq: 3}
\end{equation}
where the average is now taken over the matrix $V$.
Changing variables $\bar H_{1j} = (\frac{\pi}{B})^{\frac{1}{2}}y_j$ 
(only terms with even powers of $\bar H_{11} $ survive the 
integration over $\bar H_{11} $)
\begin{equation}
 \left \langle \mathrm{det}\bar H^{2} \mathrm{e}^{-r^2 X}
 \right \rangle_{N-1}=
 \frac{\langle \mathrm{det}V^2\rangle_{N-2}}{A^{1/2}} \left(\frac{\pi}{B}\right)^{\frac{N-2}{2}}
 \int \prod_{j=2}^{N-1} d y_j \mathrm{e}^{ -\pi \sum_{j=2}^{N-1} y_j^2}
     \left( \frac{1}{2A}+ \left(\frac{\pi}{B}\right)^{2} 
     \frac{\left \langle \left ( \sum_{j,k=2}^{N-1} y_j V^{-1}_{jk} y_k 
     \right)^2 \right \rangle_{N-2}}{\langle \mathrm{det}V^2\rangle_{N-2}} \right).
     \label{eq: 4}
\end{equation}
Denote
\begin{equation}
 \mathcal{I}^{O,2}_{N-2} = 4\pi^2 J^2\int \prod_{j=2}^{N-1} d y_j \mathrm{e}^{ -\pi \sum_{j=2}^{N-1} y_j^2}
     \frac{\left \langle \left ( \sum_{j,k=2}^{N-1} y_j V^{-1}_{jk} y_k 
     \right)^2 \right \rangle_{N-2}}{\langle \mathrm{det}V^2\rangle_{N-2}} 
     \label{eq: 4a}.
\end{equation}
Changing the order of integration and averaging in \eqref{eq: 4a}, integration over 
$y_j$ can be
done in the following way
\begin{eqnarray}
\nonumber 4\pi^2 \int \prod_{j=2}^{N-1} d y_j \mathrm{e}^{ -\pi \sum_{j=2}^{N-1} y_j^2}
     \left ( \sum_{j,k=2}^{N-1} y_j V^{-1}_{jk} y_k 
     \right)^2  = 4\pi^2\int \prod_{j=2}^{N-1} d \xi_j \mathrm{e}^{ -\pi \sum_{j=2}^{N-1} \xi_j^2}
      \sum_{j,k=2}^{N-1} \xi^2_j \xi^2_k E^{-1}_{j} E^{-1}_{j}  =\\
    = 3 \sum_{j=2}^{N-1} E^{-2}_{j}  + \sum_{j,k=2, j\neq k}^{N-1}  E^{-1}_{j} E^{-1}_{j}
    = 2 \Tr V^{-2}+\left( \Tr V^{-1}\right)^2
     \label{eq: 4b},
\end{eqnarray}
where a change of variables $z_j = O \xi_j$ such that $O^{T}V^{-1}O=\mathrm{diag}\left(
E^{-1}_{2},\ldots, E^{-1}_{N-1}\right) $ was performed. Thus
\begin{equation}
 \mathcal{I}^{O,2}_{N-2} =J^2\frac{\langle 
 \mathrm{det}V^2\left(2 \mathrm{Tr}V^{-2}+\left(\mathrm{Tr}V^{-1}\right)^2 \right)\rangle_{N-2}}
 {\langle \mathrm{det}V^2\rangle_{N-2}},
 \label{eq: 4b}
 \end{equation}
 which is precisely the form of \eqref{5c}.
The averages in \eqref{eq: 4b} contain functions of eigenvalues of $V$ -- therefore this formula is suited 
for averaging over joint probability distribution of eigenvalues for GOE. However, we can proceed 
in an easier way. Plugging in definitions of $A$ and $B$, \eqref{eq: 4} becomes
\begin{equation}
 \frac{ \left \langle \mathrm{det}\bar H^{2} \mathrm{e}^{-r^2 X}
 \right \rangle_{N-1}}{\langle \mathrm{det}V^2\rangle_{N-2} }
 = \left( \frac{1}{4 J^2 \pi } \right)^{\frac{1}{2} } \left( \frac{1}{2 J^2 \pi} \right)^{\frac{N-2}{2}}
 \left( \frac{4J^2\pi  }{1+4J^2r^2}\right)^{\frac{1}{2}}  
 \left(\frac{2J^2 \pi}{1+2J^2r^2}\right)^{\frac{N-2}{2}} 
     \left(  \frac{2J^2  }{1+4J^2r^2}+  \left(\frac{1}{1+2J^2r^2}\right)^{2} 
    J^2 \mathcal{I}^{O,2}_{N-2} \right),
     \label{eq: 3.2}
\end{equation}
where all of the normalization constants are kept. Putting $r=0$ in this formula we arrive at 
\eqref{eq: Io2n} which allows for straightforward (and \emph{exact}) calculation of $\mathcal{I}^{O,2}_N$.
Moreover, using \eqref{eq: 3.2} in \eqref{s6a}, remembering that 
$\delta\left( \chi -  2J^2 r^2\right) \propto \left(\frac{1}{\chi}\right)^{\frac{1}{2}} \delta \left(
r- \left(\frac{\chi}{2J^2}\right)^{\frac{1}{2}} \right)$ we obtain 
the fidelity susceptibility distribution for GOE \eqref{eq: 5mt}.

We finally note that the form \eqref{eq: 7mt}  of $P^O(x)$  
is such that distribution of $t=\frac{1}{x}$ is many aspects simpler:
\begin{equation}
 P(t)=\frac{1}{6}\left(1+t\right) \exp\left(-\frac{t}{2}\right),
 \label{eq: 8}
\end{equation}
which suggests that further inquires of properties of fidelity susceptibility
outside the realm of GRE could be done for $t=\frac{N}{\chi}$ variable.

 \subsection{Calculation and results for GUE}
 
Writing \eqref{s6a} for GUE - $\beta =2$, one gets
choosing $\mathbf{z} = r[1,0,0..]^T$ 
\begin{equation}
 \left \langle \mathrm{det}\bar H^{4} \mathrm{e}^{-r^2 X}
 \right \rangle_{N-1} =
 \int d\bar H_{11}\mathrm{e}^{-A \bar H_{11}^2}\prod_{j=2}^{N-1} d \bar H^R_{1j}
 d \bar H^I_{1j} \mathrm{e}^{-B \sum_{j=2}^{N-1} | \bar H_{1j}|^2}
  \mathrm{det}\bar H^{4} D^{N-2}V,
  \label{eq: 9p}
\end{equation}
with  $A = \frac{1}{2J^2} +r^2$ and $B=\frac{1}{J^2} +r^2$.
Changing variables: $\bar H_{1j} =\bar H^R_{1j}+i\bar H^I_{1j}= (\frac{\pi}{B})^{\frac{1}{2}}(x_j+ \mathrm{i}y_j)
=(\frac{\pi}{B})^{\frac{1}{2}}z_j$ 
and using the formula for determinant of block matrix one gets 
\begin{equation}
 \left \langle \mathrm{det}\bar H^{4} \mathrm{e}^{-r^2 X}
 \right \rangle_{ N-1} / \left \langle \mathrm{det}V^{4} 
 \right \rangle_{N-2}=
 \left(\frac{\pi}{A}\right)^{\frac{1}{2}} \left(\frac{\pi}{B}\right)^{N-2}
 \int \prod_{j=2}^{N-1} d x_j d y_j  \mathrm{e}^{ -\pi \sum_{j=2}^{N-1}|z_j|^2} \times 
\nonumber
\end{equation}
\begin{equation}
     \left( \frac{3}{4A^2} + 
    6\frac{1}{2 A} \left( \frac{\pi}{B} \right)^2
    \frac{\left \langle \mathrm{det}V^4 \left ( 
    \sum_{j,k=2}^{N-1} z_j V^{-1}_{jk} z^*_k 
     \right)^2 \right \rangle_{N-2} }
     {\left \langle \mathrm{det}V^{4} 
\right \rangle_{N-2}}
     +\left( \frac{\pi}{B} \right)^4 \frac{  \left \langle \mathrm{det}V^4 \left ( \sum_{j,k=2}^{N-1} z_j V^{-1}_{jk} z^*_k 
     \right)^4 \right \rangle_{N-2} }{\left \langle \mathrm{det}V^{4} 
\right \rangle_{N-2} }
     \right).
          \label{eq: 9}
\end{equation}
Denote
\begin{equation}
\mathcal{I}^{U,2}_{N-2} = J^2\pi^2 \int \prod_{j=2}^{N-1} d x_j d y_j  \mathrm{e}^{ -\pi \sum_{j=2}^{N-1}|z_j|^2}
 \frac{\left \langle \mathrm{det}V^4 \left ( 
    \sum_{j,k=2}^{N-1} z_j V^{-1}_{jk} z^*_k 
     \right)^2 \right \rangle_{N-2} }
     {\left \langle \mathrm{det}V^{4} 
\right \rangle_{N-2}}
\label{eq: I42}
\end{equation}
and
\begin{equation}
 \mathcal{I}^{U,4}_{N-2} = J^4\pi^4 \int \prod_{j=2}^{N-1} d x_j d y_j  \mathrm{e}^{ -\pi \sum_{j=2}^{N-1}|z_j|^2}
 \frac{  \left \langle \left( \mathrm{det}V^4 ( \sum_{j,k=2}^{N-1} z_j V^{-1}_{jk} z^*_k 
     \right)^4 \right \rangle_{N-2} }{\left \langle \mathrm{det}V^{4} 
\right \rangle_{N-2} }.
\label{eq: I44}
\end{equation}
Expressing $A$ and $B$ in terms of $J^2$ and $r^2$ results in
\begin{equation}
 \left \langle \mathrm{det}\bar H^{4} \mathrm{e}^{-r^2 X}
 \right \rangle_{N-1} / \left \langle \mathrm{det}V^{4} 
 \right \rangle_{N-2}=\left( \frac{1}{2 J^2 \pi } \right)^{\frac{1}{2} } \left( \frac{1}{ J^2 \pi} \right)^{N-2} \times
 \nonumber
 \end{equation}
 \begin{equation}
 \left(\frac{2J^2\pi}{1+2J^2r^2}\right)^{\frac{1}{2}} \left(\frac{J^2\pi}{1+J^2r^2}\right)^{N-2}
     \left[ \frac{3}{4} \left( \frac{2J^2}{1+2J^2r^2}\right)^{2} + 
    3  \frac{2}{1+2J^2r^2} \left( \frac{J^2}{1+J^2r^2} \right)^2
    \mathcal{I}^{U,2}_{N-2}
     +\left( \frac{J}{1+J^2r^2} \right)^4 \mathcal{I}^{U,4}_{N-2}
     \right].
          \label{eq: 99}
\end{equation}
First of all, this equation used in \eqref{s6a} implies the form of the fidelity susceptibility
distribution for GUE \eqref{eq: 10}. Moreover, taking $r=0$ in \eqref{eq: 99} and 
using expression for the fourth moment of determinant of GUE matrix from \cite{Mehta98, Cicuta00}
we get that
\begin{equation}
 \left \langle \mathrm{det}\bar H^{4}
 \right \rangle_{N-1} / \left \langle \mathrm{det}V^{4} 
 \right \rangle_{N-2}=
\begin{cases}
J^4 (N^2-1), \quad  \quad \quad N \quad \mathrm{even}, \\ 
J^4 (N^2+2N) \quad  \quad N \quad \mathrm{odd}.
\end{cases}
 \label{eq: 99c}
\end{equation}
which, together with the
exact result for $\mathcal{I}^{U,2}_N$ obtained below (equations \eqref{eq: 8.8}, \eqref{eq: 8.9}) 
is equivalent to \eqref{iu4a}. To complete the derivation of fidelity susceptibility  
we need to address the task of calculating $\mathcal{I}^{U,2}_{N}$ to which we turn now.


Let us start by expressing $\mathcal{I}^{U,2}_N$ in terms of invariants ($H$ is now
$N\times N$ GUE matrix),
\begin{equation}
 \mathcal{I}^{U,2}_N =\frac{ J^2 \pi^2}{\left \langle \mathrm{det}H^{4} 
\right \rangle_{N}} \left \langle \mathrm{det}H^4
\int \prod_{j=1}^{N} d x_j d y_j  \mathrm{e}^{ -\pi \sum_{j=1}^{N}|z_j|^2}
  \left ( 
    \sum_{j,k=1}^{N} z_j H^{-1}_{jk} z^*_k 
     \right)^2 \right \rangle_{N} \equiv
     \frac{ J^2}{\left \langle \mathrm{det}H^{4} 
\right \rangle_{N}} \left \langle \mathrm{det}H^4 I^{U,2}_N \right \rangle_{N}.
\label{eq: 5.1s}
\end{equation}
Substituting $z_i = U \xi_i$ with $U$ such that $UH^{-1}U^{\dag} = \mathrm{diag}
\left( E^{-1}_1,..., E^{-1}_N\right)$ and then putting
$\xi_i=r_i \mathrm{e}^{\mathrm{i}\phi_i}$  one gets
\begin{equation}
 I^{U,2}_N = \pi^2 \int \prod_{j=1}^{N} d r_j d \phi_j r_j
 \mathrm{e}^{ -\pi \sum_{j=1}^{N}r_j^2}
 \sum_{j,l}r_j^2 r_l^2 E_j^{-1} E_l^{-1}.
 \label{eq: 5.2}
\end{equation}
One can integrate over the phases $\phi_j$, resulting in a factor $(2\pi)^N$ 
which cancels out with $1/(2\pi)^N$  arising in substitution 
$t_i=\pi r_i^2$ so that the integral becomes
\begin{equation}
 I^{U,2}_N = \int \prod_{j} d t_j 
 \mathrm{e}^{ - \sum_{j}t_j}
 \sum_{j}t_j t_l E_j^{-1} E_l^{-1}=
 m_2\sum_{j} E_j^{-2} + m_1^2 \sum_{j\neq l}E_j^{-1} E_l^{-1},
 \label{eq: 5.3}
\end{equation}
where $m_2$ and $m_1$ are the second and the first moments of $\mathrm{e}^{-t}$ distribution.
Using \eqref{eq: 5.3} in \eqref{eq: 5.1s}, remembering that $m_2=2$ and $m_1=1$
one obtains the following
expression 
\begin{equation}
 \mathcal{I}^{U,2}_N =
     \frac{ J^2}{\left \langle \mathrm{det}H^{4} 
\right \rangle_{N}} \left \langle \mathrm{det}H^4 
\left(\mathrm{Tr}H^{-2} + ( \mathrm{Tr}{H^{-1}} )^2
\right) \right \rangle_{N},
\label{eq: 5.1}
\end{equation}
demonstrating validity of \eqref{eq: I42a}.

\subsection{The generating function}

Consider the generating function \eqref{eq: 7.1mt}
\begin{equation}
 Z_N(j_1,j_2) = \left \langle \det H^2 \det(H-j_1) \det(H-j_2) \right \rangle_{N}.
 \label{eq: 7.1}
\end{equation}
Using the equality 
\begin{equation}
 \frac{\partial}{\partial j} \det(H-j) = 
 \frac{\partial}{\partial j} \prod_{k=1}^N (E_k - j)=
 -\sum_l \frac{\prod_{k=1}^N (E_k - j)}{E_l-j} = -\det(H-j) \mathrm{Tr}(H-j)^{-1}
 \label{eq: 7.2}
\end{equation}
we verify that \eqref{eq: 7.5mt} indeed holds.
Moreover, as a side product one gets
\begin{equation}
 \frac{\left \langle \det H \right \rangle_{N+1}}{\left \langle \det H \right \rangle_{N}}
 =\frac{Z_{N+1}(0,0)}{Z_N(0,0)}
 \stackrel{\lim_{N\rightarrow\infty}}{=}J^4\mathcal{I}^{U,4}_N    
\label{eq: 7.6}
\end{equation}
which can be used as a validation of our calculation by comparison of the result with \eqref{eq: 99c}.

\subsection{Calculation of generating function}

Formulas best suited for our task of finding $Z(j_1,j_2)$ 
are worked out in \cite{Strahov03}:
\begin{equation}
 \left \langle \prod_{j=1}^K \det(H - \lambda_j)\det(H - \mu_j) \right \rangle_N=
 \frac{C_{N,K}}{\Delta(\lambda_1,...,\lambda_K)
 \Delta(\mu_1,...,\mu_K)}
 \det\left[W_{N+K}(\lambda_i, \mu_j) \right]_{i,j=1,...,K},
\label{eq: 8.1} 
\end{equation}
where $\Delta(\lambda_1,...,\lambda_K)$ is Vandermonde determinant and
the kernel $W_{N+K}$ reads 
\begin{equation}
W_{N+K}(\lambda, \mu) 
=
\frac{1}{\lambda-\mu} \left[ \Pi_{N+K}(\lambda) \Pi_{N+K-1}(\mu) - 
\Pi_{N+K}(\mu) \Pi_{N+K-1}(\lambda)
\right]
\label{eq: 8.2c} 
\end{equation}
where $\Pi_{M}(\lambda)$ are monic polynomials orthogonal 
with respect to a measure
$\mathrm{e}^{-V(x)} \mathrm{d}x$ and $C_{N,K}$ is a constant. For the GUE case $V(x)=\frac{1}{2J^2}x^2$.
Using the equations \eqref{eq: 8.3}, \eqref{eq: 8.2} -- we obtain the following closed analytical expression 
for the generating function $Z(j_1,j_2)$
 \begin{equation}
 \nonumber
 Z_N(j_1, j_2) =  -2C_{N,2} \frac{ \pi J^{4 N +6}  }{j_1^2 j_2^2\Gamma \left(-\frac{N+1}{2}\right)^2}
   H_{N +1}\left(\frac{j_1}{ \sqrt{2J^2} }\right) H_{N +1}\left(\frac{j_2}{ \sqrt{2J^2} }\right)
  +
 \end{equation}
 \begin{equation} + 
  C_{N,2} \frac{ \pi J^{4 N +6}  (N +1) }{j_1 j_2(j_1-j_2) J  \sqrt{2} \Gamma \left(-\frac{N+1}{2}\right) \Gamma \left(\frac{1-N }{2}\right)}\left(
     H_{N +1}\left(\frac{j_1}{ \sqrt{2J^2} }\right) H_{N +2}\left(\frac{j_2}{ \sqrt{2J^2} }\right)
  -    H_{N +2}\left(\frac{j_1}{ \sqrt{2J^2} }\right) H_{N +1}\left(\frac{j_2}{ \sqrt{2J^2} }\right)\right)
\label{eq: 8.5}
\end{equation}
for even $N$. Expression for odd $N$ can be analogously derived.

It is interesting to note that von Oppen, during his calculation of distribution of curvatures
for GUE \cite{vonOppen94} calculated
\begin{equation}
\left \langle \det H^3(\det H - j_2)\right \rangle_{N}
\sim
\frac{\sin\left( \sqrt{\frac{N}{J^2}} j_2 \right) -\sqrt{\frac{N}{J^2}}j_2 
\cos\left( \sqrt{\frac{N}{J^2}} E_2\right)}{(\sqrt{\frac{N}{J^2}}E_2)^3}
\label{q2d} 
\end{equation}
using technique of supersymmetric integrals (for a pedagogical introduction of this technique see \cite{Haake}).
The formula for $Z(j_1,j_2)$ derived by us is an extension of the above expression--
one can show that in the limit $\lim_{j_1 \rightarrow 0}Z_N(j_1,j_2)$ one recovers the von Oppen's formula \eqref{q2d}
for $N\gg1$.
Calculating the derivatives one readily obtains:
\begin{equation}
\frac{\partial^2}{\partial j_1 \partial j_2} Z_N(0,0) = 
\begin{cases}
\frac{1}{15}N, \quad  \quad \quad \quad \,\,\, N \quad \mathrm{even}, \\ 
\frac{1}{15}(N+4),\quad  \quad N \quad \mathrm{odd}.
\end{cases}
\label{eq: 8.7}
\end{equation}
and 
\begin{equation}
\frac{\partial^2}{\partial j_2^2 } Z_N(0,0) = 
\begin{cases}
-\frac{1}{5}N, \quad  \quad  \quad \quad \,\,\, N \quad \mathrm{even}, \\ 
-\frac{1}{5}(N-1),\quad  \quad N \quad \mathrm{odd}.
\end{cases}
\label{eq: 8.8}
\end{equation}
which via \eqref{eq: 7.5mt} implies that
\begin{equation}
 \mathcal{I}^{U,2}_N=
\begin{cases}
\frac{1}{3}N, \quad  \quad \quad \quad \,\,\, N \quad \mathrm{even}, \\ 
\frac{1}{3}(N+1), \quad  \quad N \quad \mathrm{odd}.
\end{cases}
\label{eq: 8.9}
\end{equation}

%

\end{widetext}

\end{document}